\documentclass[twocolumn]{aastex631}

\usepackage{subfigure}
\usepackage{makecell}

\usepackage{changepage}
\usepackage{booktabs}
\usepackage{comment}
\usepackage{threeparttable}
\usepackage{amssymb}
\usepackage{graphicx}
\usepackage{CJK}
\usepackage{diagbox}
\usepackage{multirow}
\usepackage[figureright]{rotating}
\usepackage{tablefootnote}
\usepackage{amsmath}
\usepackage{threeparttable}
\usepackage{xspace}

\newcommand{\Ha}{\textrm{H}\ensuremath{\alpha}\xspace}
\newcommand{\Hb}{\textrm{H}\ensuremath{\beta}\xspace}
\newcommand{\Hg}{\textrm{H}\ensuremath{\gamma}\xspace}

\newcommand{\HII}{\textrm{H}\textsc{ii}\xspace}
\newcommand{\HI}{\textrm{H}\textsc{i}\xspace}

\newcommand{\OI}{[\textrm{O}~\textsc{i}]\xspace}
\newcommand{\OII}{[\textrm{O}~\textsc{ii}]\xspace}
\newcommand{\OIII}{[\textrm{O}~\textsc{iii}]\xspace}

\newcommand{\NII}{[\textrm{N}~\textsc{ii}]\xspace}
\newcommand{\SII}{[\textrm{S}~\textsc{ii}]\xspace}

\begin{document}
\begin{CJK*}{UTF8}{gbsn}

\title{MAMMOTH-MOSFIRE: Environmental Effects on Galaxy Interstellar Medium at $z\sim2$}

\correspondingauthor{Xin Wang}
\email{xwang@ucas.ac.cn}


\author[0009-0004-7133-9375]{Hang Zhou}
\affil{School of Astronomy and Space Science, University of Chinese Academy of Sciences (UCAS), Beijing 100049, China}

\author[0000-0002-9373-3865]{Xin Wang}
\affil{School of Astronomy and Space Science, University of Chinese Academy of Sciences (UCAS), Beijing 100049, China}
\affil{National Astronomical Observatories, Chinese Academy of Sciences, Beijing 100101, China}
\affil{Institute for Frontiers in Astronomy and Astrophysics, Beijing Normal University,  Beijing 102206, China}

\author[0000-0001-6919-1237]{Matthew A. Malkan}
\affiliation{Department of Physics and Astronomy, University of California, Los Angeles, 430 Portola Plaza, Los Angeles, CA 90095, USA}

\author[0000-0002-8460-0390]{Tommaso Treu}
\affiliation{Department of Physics and Astronomy, University of California, Los Angeles, 430 Portola Plaza, Los Angeles, CA 90095, USA}

\author[0000-0002-0663-814X]{Yiming Yang}
\affil{National Astronomical Observatories, Chinese Academy of Sciences, Beijing 100101, China}
\affil{School of Astronomy and Space Science, University of Chinese Academy of Sciences (UCAS), Beijing 100049, China}


\author[0000-0001-8467-6478]{Zheng Cai}
\affiliation{Department of Astronomy, Tsinghua University, Beijing 100084, China}

\author[0000-0003-3310-0131]{Xiaohui Fan}
\affiliation{Steward Observatory, University of Arizona, 933 North Cherry Ave., Tucson, AZ 85721, USA}

\author[0000-0002-5815-2387]{Mengting Ju}
\affiliation{School of Astronomy and Space Science, University of Chinese Academy of Sciences (UCAS), Beijing 100049, China}

\author[0000-0002-3264-819X]{Dong Dong Shi}
\affiliation{Center for Fundamental Physics, School of Mechanics and Optoelectronic Physics, Anhui University of Science and Technology,
168 Taifeng Street, Huainan 232001, China}


\author[0000-0002-8630-6435]{Anahita Alavi}
\affiliation{Infrared Processing and Analysis Center, Caltech, 1200 E. California Blvd., Pasadena, CA 91125, USA}

\author[0000-0002-1620-0897]{Fuyan Bian}
\affiliation{European Southern Observatory, Alonso de Cordova 3107, Casilla 19001, Vitacura, Santiago 19, Chile}

\author[0000-0001-6482-3020]{James Colbert}
\affiliation{Infrared Processing and Analysis Center, Caltech, 1200 E. California Blvd., Pasadena, CA 91125, USA}

\author[0000-0002-6586-4446]{Alaina L. Henry}
\affiliation{Space Telescope Science Institute, 3700 San Martin Dr., Baltimore, MD, 21218, USA}

\author[0000-0003-4813-8482]{Sijia Li}
\affil{School of Astronomy and Space Science, University of Chinese Academy of Sciences (UCAS), Beijing 100049, China}
\affil{Department of Astronomy, Xiamen University, Xiamen, Fujian 361005, China}

\author[0000-0001-5951-459X]{Zihao Li}
\affiliation{Cosmic Dawn Center (DAWN), Denmark}
\affiliation{Niels Bohr Institute, University of Copenhagen, Jagtvej 128, DK2200 Copenhagen N, Denmark}
\affiliation{Department of Astronomy, Tsinghua University, Beijing 100084, China}

\author[0000-0002-7064-5424]{Harry I. Teplitz}
\affiliation{Infrared Processing and Analysis Center, Caltech, 1200 E. California Blvd., Pasadena, CA 91125, USA}

\author[0000-0003-1718-6481]{Hu Zhan}
\affil{National Astronomical Observatories, Chinese Academy of Sciences, Beijing 100101, China}
\affil{The Kavli Institute for Astronomy and Astrophysics, Peking University, Beijing 100871, China}

\author[0000-0003-3728-9912]{Xian~Zhong Zheng}
\affil{Purple Mountain Observatory, Chinese Academy of Sciences, 10 Yuan Hua Road, Nanjing 210023, China}

\author{Zheng Zheng}
\affil{National Astronomical Observatories, Chinese Academy of Sciences, Beijing 100101, China}

\author[0000-0003-0401-3688]{Yifei Jin}
\affiliation{Westlake University, 600 Dunyu Road, Xihu District, Hangzhou, Zhejiang 310030 PR China}

\begin{abstract}

The MAMMOTH-MOSFIRE program is a deep Keck/MOSFIRE K-band spectroscopic follow-up of emission-line galaxies identified in the MAMMOTH-Grism HST/WFC3 G141 slitless spectroscopic survey, targeting the core regions of three most massive galaxy protoclusters at cosmic noon.
To introduce this program, we present a comprehensive analysis of the $\NII\lambda$6584, $\SII\lambda\lambda$6717,6731, and $\OI\lambda$6300 BPT diagnostics for a unique sample of 43 protocluster member galaxies at $z\sim2$, investigating how the overdense environment influences their interstellar medium ionization conditions. 
We find that, similar to their field counterparts at $z\sim2$, protocluster galaxies exhibit a systematic offset in the \NII BPT diagram relative to the local star-forming sequence, but no such offset in the \SII BPT diagram.
Notably, we detect significantly elevated \OI/\Ha ratios, which can be well reproduced by photoionization models incorporating both \HII regions and low-velocity shocks ($v$ $\sim$ 200 km s$^{-1}$).
We caution that neglecting shock excitation can bias abundance measurements, leading to an overestimation of nitrogen enrichment. This provides a potential explanation for the long-standing puzzle of enhanced $\NII/\Ha$ ratios observed in $z\sim2$ galaxies.
We further suggest that these shocks are likely environmentally driven, e.g., by ram-pressure stripping or tidal interactions, which requires future confirmation through direct observations of features such as stripped gas tails.

\end{abstract}


\section{Introduction} \label{sec::intro}

Understanding galaxy formation and evolution is a pivotal task of modern astronomy. Galaxy protoclusters at $z\sim$ 2-3 provide a unique laboratory to observe how galaxies evolve in dense environments. Galaxy protoclusters also provide an opportunity to observe the galaxy formation processes under the conditions found in the early universe \citep{chiangGalaxyProtoclustersDrivers2017}. These studies could reveal the detailed physical mechanisms of star formation and galactic feedback in the early universe under the rapid assembly of the large-scale structures \citep{Overzier_2016,Shimakawa18}. 



One such crucial interstellar medium (ISM) property is the average electron density ($n_e$)
\citep[e.g.,][]
{kewley_electron,Hainline_2009,Lehnert_2009}, which has been measured in $z\sim$2 high redshift galaxies \citep{kewley_electron,Sanders_2016,Reddy_2023}. Past research indicates that the emission line ratios of high-redshift galaxies are offset from those of local galaxies. 
The Baldwin, Philips, $\&$ Terlevich (BPT) diagram \citep{Baldwin1981, Veilleux1987},  plotting the ratios of $\OIII\lambda$5007/\Hb versus $\NII\lambda$6584/\Ha \citep{Kewley_2001, Kauffmann2003a}, is commonly used for distinguishing different types of galaxies, such as star-forming galaxies and active galactic nuclei (AGN) \citep[e.g.,][]{Dopita_2013,Cameron_2020,Dopita_2016,garg2022,Pagotto_2021}.
The physical properties of star-forming regions in these high-redshift galaxies that drive the observed changes in emission-line ratios remain unclear. Possible contributors include higher ionization parameters and/or electron densities \citep[e.g.,][]{Brinchmann2008,Kewley_2015,Masters_2014}, harder ionizing radiation fields \citep{Steidel_2014}, contributions from shocks \citep{Newman_2014}, and a variation in N/O ratio \citep{Shapley15}.  

To answer these questions, we compile a large sample of emission-line galaxies identified in the centers of three most massive galaxy protoclusters known at $z\sim2.2$. We combine  the Hubble Space Telescope (HST) wide-field camera 3 (WFC3) near-infrared G141 slitless spectroscopy covering the J and H bands, and the Multi-Object Spectrometer For Infra-Red Exploration (MOSFIRE) on the Keck-I telescope covering the K-band, we secure the full suite of the rest-frame optical diagnostic emission lines of $\OII\lambda\lambda$3727,3729, $\OIII\lambda$5007, \Hb, \OI, \Ha, $\NII\lambda$6584, $\SII\lambda\lambda$6717,6731 for this unique sample of protocluster member galaxies at the cosmic noon \citep{Melanie16}.

The structure of the paper is as follows: In Section~\ref{sec:OBSERVATIONS}, we briefly introduce the data reduction process, including observations, sample selection, and spectral stacking. We present the detailed measurements in Section~\ref{sec:measure}.
Finally, a discussion of our findings is provided in Section~\ref{sec:discussion}.

We adopt the following abbreviations to refer to commonly used emission line ratios:
\begin{equation}\label{eq::N2}
\mathrm{N2} = {\NII\lambda6584}/{\mathrm{H}\alpha}
\end{equation}
\begin{equation}\label{eq::O1}
\mathrm{O1} = {\OI\lambda6300}/{\mathrm{H}\alpha}
\end{equation}
\begin{equation}\label{eq::S2}
\mathrm{S2} = {\SII\lambda\lambda6717,6731}/{\mathrm{H}\alpha}
\end{equation}
\begin{equation}\label{eq::O32}
\mathrm{O32} = {\OIII\lambda\lambda4959,5007}/{\OII\lambda\lambda3727,3729}
\end{equation}
\begin{equation}\label{eq::R23}
\mathrm{R23} = ({{\OIII\lambda\lambda4959,5007}+\OII\lambda\lambda3727,3729})/{\mathrm{H}\beta}
\end{equation}
We assume a standard $\Lambda$CDM cosmology with parameters of ($\Omega_m$, $\Omega_\Lambda$, H$_0$) = (0.3, 0.7, 70 km s$^{-1}$ Mpc$^{-1}$).
Throughout the paper, we abbreviate the forbidden lines with 
$\OI\lambda6300=  \OI$,
$\NII\lambda6584= \NII$,
$\SII\lambda\lambda6717,6731= \SII$, 
$\OIII\lambda5007= \OIII$, 
if presented without wavelength values.

\begin{deluxetable*}{lcccccc}
    \label{table:field}
    \tablecolumns{7}
    \tablewidth{0pt}
    \tablecaption{Properties of the three galaxy protocluster fields focused upon in this work.}

\tablehead{
    Protocluster & RA  & DEC & $z$ & $\delta_{\rm g}$\tablenotemark{a} & $M^{\rm z=0}_{\rm tot}$\tablenotemark{b}  & $N_{\rm gal}$\tablenotemark{c} \\
    Field  & [deg.]  & [deg.] & & &$10^{15}$M$_\odot$  & 
}
\startdata
    BOSS1244 &  190.9  & 35.9 &  2.24 & 22.9 $\pm$ 4.9 & 1.6 $\pm$ 0.2 & 27\\ 
    BOSS1441  &  220.3 & 40.0 & 2.32 & 10.8 $\pm$ 2.8 & 1.0 $\pm$ 0.2 & 5 \\ 
    BOSS1542  &  235.7 & 38.9 & 2.24 & 20.5 $\pm$ 3.9 &  1.4 $\pm$ 0.2 & 11  
\enddata
    \tablenotetext{a}{The galaxy overdensity $\delta_{\rm g}=\Sigma_{\rm cluster}/\Sigma_{\rm field}-1$, where $\Sigma_{\rm cluster/field}$ correspond to the galaxy surface densities measured in the protocluster/blank fields. These surface densities are averaged over an area of 15 cMpc$^{2}$. For BOSS1244 and BOSS1542, $\delta_{\rm g}$ is estimated based upon \Ha emitters \citep{Shi_2021}, whereas for BOSS1441, $\delta_{\rm g}$ comes from Ly$\alpha$ emitters \citep{Cai_2017}.}
    \tablenotetext{b}{The present-day total enclosed mass derived from $\delta_{\rm gal}$ for the three protoclusters \citep{Cai_2017,Shi_2021}.}
    \tablenotetext{c}{Number of \Ha emitters confirmed by our Keck MOSFIRE spectroscopy presented in this work in each protocluster field.}
\end{deluxetable*}

\begin{deluxetable*}{lccccccccccc}
\vspace{-.8cm}
\label{table:keck}
\tablecolumns{8}
\tablewidth{0pt}
\tablecaption{Details of our Keck MOSFIRE spectroscopy of the protocluster member galaxies in the BOSS1244, BOSS1441 and BOSS1542 fields. The observations were taken in 3 full nights from 2022A\_U016 and 2023A\_U139 (PI M. Malkan).}
\tablehead{
    Field/Slit Mask & $N_{\rm obj}$\tablenotemark{a} &  RA  & DEC & Position Angle & Filter & Number of Repeats\tablenotemark{b} & Observation Date  \\
      &  &(J2000.0)  & (J2000.0) & (deg) &  & (s) & 
}
\startdata
    BOSS1244/mask1.3.3 & 6/22 & 12$^{h}$43$^{m}$36$^{s}$.38 & 35$^{d}$52$^{m}$45$^{s}$.73 & 331  & K & 10 & 04/15/2022  \\
    BOSS1244/mask2.2.3 & 7/23 & 12$^{h}$43$^{m}$29$^{s}$.44  & 35$^{d}$56$^{m}$02$^{s}$.33 & 133  & K & 11 & 04/15/2022  \\
    BOSS1244/mask5.3  & 9/26 &  12$^{h}$43$^{m}$31$^{s}$.99 & 35$^{d}$54$^{m}$18$^{s}$.80 & 11.5  & K & 11 & 05/06/2023 \\
    BOSS1244/mask6.6 &  7/18 &  12$^{h}$43$^{m}$24$^{s}$.51 & 35$^{d}$55$^{m}$14$^{s}$.84 & 344   & K & 8  & 05/23/2023 \\
    BOSS1441/mask2.6 & 5/22 &  14$^{h}$41$^{m}$30$^{s}$.81 & 40$^{d}$02$^{m}$44$^{s}$.35 & 70     & K & 10 & 05/06/2023 \\
    BOSS1542/mask1.3.1 & 5/22 &  15$^{h}$42$^{m}$46$^{s}$.05& 38$^{d}$58$^{m}$51$^{s}$.16 & 354   & K & 13 & 04/15/2022 \\
    BOSS1542/mask2.2 & 3/22 &  15$^{h}$42$^{m}$50$^{s}$.17 & 38$^{d}$51$^{m}$55$^{s}$.13 & 355      & K & 10 & 05/23/2023 \\
    BOSS1542/mask3.3 & 1/17 &  15$^{h}$42$^{m}$50$^{s}$.86 & 38$^{d}$51$^{m}$55$^{s}$.44 & 309     & K & 13 & 05/23/2023 
\enddata
    \tablenotetext{a}{The number on the right refers to the total number of objects put on each mask, whereas the number on the left corresponds to the number of \Ha emitters analyzed in this work.}
    \tablenotetext{b}{One repeat of exposure set is comprised of one standard ABA'B' dither pattern with each exposure having an integration time of 180 seconds, excluding overheads. This is the recommended exposure strategy for K filter spectroscopy, with the default slit width of 0.7" and the MCDS16 sampling mode.}
\vspace{-.5cm}
\end{deluxetable*}

\section{Data and Analysis}
\label{sec:OBSERVATIONS}

In this work, we focus on the ISM ionization mechanisms in overdense galaxy environments. Specifically, we examine massive galaxy protoclusters in three fields—BOSS1244, BOSS1441, and BOSS1542—identified using the Mapping the Most Massive Overdensity Through Hydrogen (MAMMOTH) technique \citep{Cai_2017,Shi_2021}. 

We show the properties and statistics of our target protocluster fields in Table~\ref{table:field}. The redshifts of the three massive protocluster fields are $z \sim 2.2-2.3$. The galaxy overdensity, $\delta_g$, is defined as $\rm \delta_g = \frac{\Sigma_{cluster}}{\Sigma_{field}}-1$, with $\rm \frac{\Sigma_{cluster}}{\Sigma_{field}}$ corresponding to the \Ha emitters (HAEs) or Ly$\alpha$ emitters (LAEs) number counts per arcmin$^2$ in the protocluster/blank fields, respectively, measured on a comoving scale of 15 cMpc. The total mass refers to the enclosed mass expected at the present day based on the value of $\delta_g$.

\subsection{HST Observations}

\begin{figure*}[t]
\centering  
\subfigure{
\includegraphics[width=0.9\linewidth]{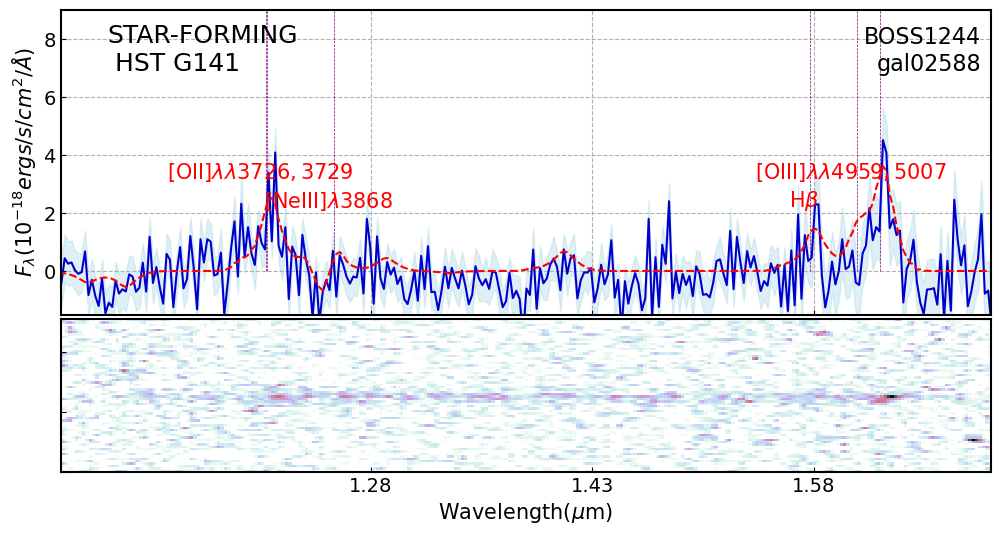}}
\caption{Flux-calibrated G141 grism spectra for object gal02588, as listed in Table~\ref{table::measure}. The spectrum is from HST observations and has been processed using GRIZLI software. The 1D spectrum is displayed as a blue line, while the Gaussian fit is overplotted in red. The uncertainty is illustrated by the light blue shaded area. The bottom panel presents the corresponding 2D spectrum. 
}
\label{fig:show_galaxy_grizli}
\end{figure*}

In HST cycle 28, we conduct the MAMMOTH-Grism slitless spectroscopic survey (GO-16276, PI: X. Wang), to obtain deep Wide-Field Camera 3 (WFC3) G141 exposures in these three massive protocluster fields \citep{Wang_2022}. The core region of each protocluster is covered by five individual WFC3 pointings, each at 3-orbit depth of G141 slitless spectroscopy. The corresponding HST footprints are shown in Figure~\ref{fig:footprint}.
These observations are designed to capture key emission lines from galaxies at z$\sim$2.2, notably \OIII, \Hb, \Hg and \OII, efficiently confirming protocluster membership. The G141 spectrum of one example galaxy is shown in Figure~\ref{fig:show_galaxy_grizli}.
Using these G141 data, we identify $\sim$50 emission-line galaxies in each field as protocluster members. 

Using images from HST and ground-based telescopes, we estimate the stellar masses of these protocluster member galaxies with BAGPIPES \citep{bagpipes}, adopting the initial mass function from \cite{Chabrier_2003} and the extinction law from \cite{Calzetti_2000}.
Our results of mass are shown in the Table~\ref{table::measure}. Further information will be forthcoming in Yang et al. (2025, in prep).

\subsection{Target Selection for Keck Spectroscopic Follow-up}

We follow our previous work to analyze the HST WFC3/G141 slitless spectroscopic data using the GRIZLI\footnote{\url{https://github.com/gbrammer/grizli/}} software \citep{wangDiscoveryStronglyInverted2019,wangCensusSubkiloparsecResolution2020,Wang_2022,wangEarlyResultsGLASSJWST2022}.
GRIZLI performs forward modeling of paired grism and pre-imaging exposures, fitting the optimally extracted one-dimensional (1D) spectra with linear combinations of spectral templates to determine the best-fit grism redshifts.
This procedure is particularly robust for our targets at the protocluster redshifts ($z\sim2.2-2.3$), owing to the presence of strong nebular emission lines (\OIII, \Hb, \Hg and \OII). We therefore select galaxies with secure grism redshifts that are spectroscopically confirmed members of the protoclusters. 
Among the confirmed protocluster members, we further prioritize galaxies with significant emission-line detections --- specifically signal-to-noise ratio (SNR) $\gtrsim$ 2 in \OIII, \Hb, and \OII --- by assigning higher target weights during MOSFIRE slit-mask design (see Sect.~\ref{subsec:reduction}).

\subsection{Keck Observations and Reduction}
\label{subsec:reduction}

\begin{figure*}[!htb]
\centering  
\subfigure{
\includegraphics[width=0.3\linewidth]{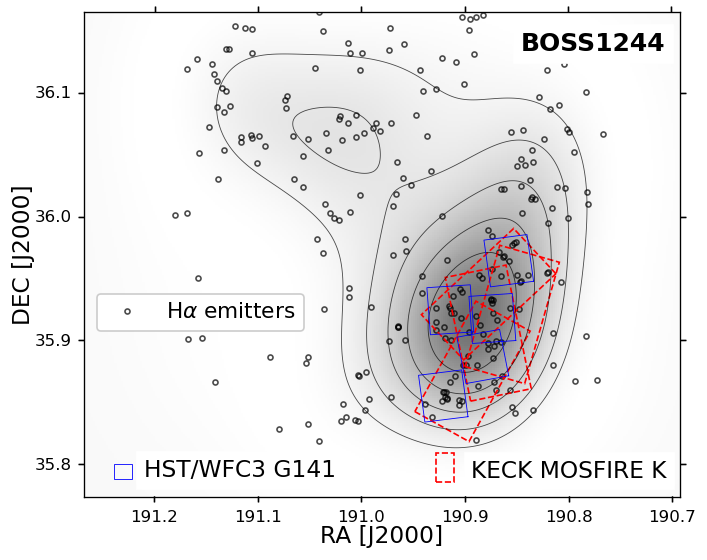}}
\subfigure{
\includegraphics[width=0.305\linewidth]{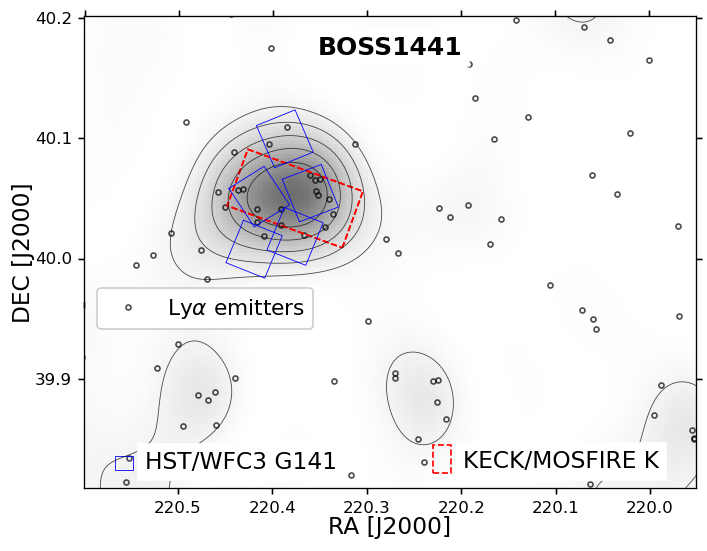}}
\subfigure{
\includegraphics[width=0.3\linewidth]{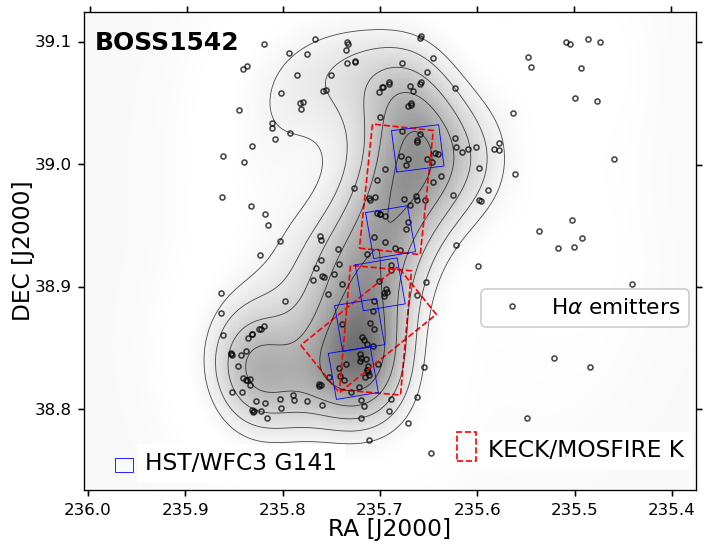}}
\caption{
Density maps of the three massive galaxy protoclusters at $z\approx2.2$ presented in this work.
In all three panels, the gray thin contours mark galaxy overdensity ($\delta_{\rm g}$) in steps of 0.2\,arcmin$^{-2}$, with the inner density peaks reaching $\sim$~2\,arcmin$^{-2}$. 
For the BOSS1244 and BOSS1542 protocluster fields, $\delta_{\rm g}$ is estimated based upon the surface density of the \Ha emitters (HAEs) at the protocluster redshift, identified from ground-based broad+narrow band imaging, whereas for the BOSS1441 field, $\delta_{\rm g}$ is from the Ly$\alpha$ emitters (LAEs) \citep{Cai_2017,Shi_2021}.
The blue squares mark the locations of the HST WFC3/G141 pointings of the MAMMOTH-Grism survey \citep[GO-16276, PI Wang,][]{Wang_2022}, and the red rectangles (each with a 3'$\times$6' FoV) show the masks of our Keck/MOSFIRE K-band followup spectroscopy (2022A\_U016, 2023A\_U139, PI Malkan).
The general properties of our three protocluster fields are given in Table~\ref{table:field}, and the detailed information about our Keck observations is shown in Table~\ref{table:keck}.}
\label{fig:footprint}
\end{figure*}

To acquire K-band spectroscopy (1.91 $-$ 2.42 $\mu$m) of the confirmed protocluster member galaxies, we carry out the MAMMOTH-MOSFIRE program using the MOSFIRE instrument on the Keck I telescope under observing programs 2022A\_U016 and 2023A\_U139 (PI: M. Malkan).
These MOSFIRE K-band data cover multiple key rest-frame optical nebular emission lines (e.g. \Ha, \NII, \SII, \OI) of our protocluster member galaxies at $z\sim2.2-2.3$, highly complementary to those secured by our HST G141 data.
In Table~\ref{table:keck}, we list the detailed information about Keck observations.
To trace our target galaxies, we use MOSFIRE Automatic GUI-based Mask Application (MAGMA) software, which is designed to configure observation masks and execute the corresponding slit configurations at the telescope. The spectroscopic observation regions, covered by Keck MOSFIRE masks, are indicated by red rectangles ($3'\times6'$) in each panel of Figure~\ref{fig:footprint} within the central density peaks of the three massive protocluster fields. The background maps and contours represent the density distribution, while the HAE candidates, identified through ground-based broad+narrow band imaging, are marked by open black circles.

The MOSFIRE data reduction was performed using the PYPEIT package \citep{pypeit:zenodo}. For wavelength calibration, PYPEIT software uses the OH atmospheric lines instead of arc lamp exposures. The 1D spectra from each exposure were then co-added to enhance the SNR. The sensitivity function was built by standard star to calibrate the 1D spectra of galaxies.

\subsection{Emission Line Fluxes}
\label{sec:fitting}
Emission line ratios are commonly used to study the ionization mechanisms of the interstellar medium \citep{Baldwin1981}. Measuring these ratios requires accurate determination of emission line fluxes, which in turn depends on identifying the observational wavelengths of the emission lines. This identification relies on reliable redshift measurements for each galaxy. While the protoclusters (BOSS1244, BOSS1441, BOSS1542) have a general redshift of $z \sim 2.2-2.3$, the exact redshifts of individual galaxies remain undetermined. Therefore, it is essential to first determine the redshifts of the 43 galaxies.

We determine the redshifts of the galaxies using the peak pixel wavelength of the \Ha and \NII emission lines. Each emission line is modeled using a Gaussian function constructed with the Python package LMFIT \citep{lmfit}. 
To ensure reliability, we manually inspect the \Ha peaks of all galaxies to exclude interference from sky lines or other emission lines. The final redshifts of the galaxies are listed in Table~\ref{table::measure}.

With the redshifts determined, the specific wavelengths of the \OI, \NII doublet, \Ha, and \SII doublet emission lines are calculated. The fluxes and uncertainties of these lines are measured by fitting a Gaussian function to each emission line. We performed 1000 iterations of Gaussian fitting based on the uncertainties of the spectra, obtaining both the Gaussian model and the associated errors for each emission line.
The median SNR of the \Ha emission line fluxes is 31.8, while the uncertainties on emission line flux are based on the 68th percentile confidence intervals, representing a typical observational noise level. 

We use the Python package SPECUTILS \citep{Specutils} to fit the continuum of the entire spectrum with a linear model. The emission line component is isolated by subtracting the continuum model from the observed spectrum. The uncertainties of the emission line components are identical to these of the observed spectrum.

For emission-line fitting, we employ six three-parameter Gaussian functions, with their centers corresponding to the wavelengths of the pixels with the peak for the six emission lines. During the fitting process, we constrained the Gaussian widths ($\sigma$) of the doublets to be consistent. For example, $\NII\lambda$6548 and $\NII\lambda$6584 share the same $\sigma$, as do $\SII\lambda$6717 and $\SII\lambda$6731. We perform 1000 iterations for each Gaussian fit based on the spectral errors. The mean value of the integrated flux from these iterations was taken as the emission line flux, and the standard deviation was used as its uncertainty. The line ratios log(\OIII/\Hb), log(\NII/\Ha), log(\SII/\Ha), and log(\OI/\Ha) are listed in Table~\ref{table::measure}.


We measured the FWHM of the \Ha emission lines in galaxy spectra and found the distribution value of $188\pm55{\rm km s^{-1}}$. When analyzing G141-band spectra with GRIZLI, a key advantage is that GRIZLI accounts for morphological broadening, which fixes the grism FWHM at about 1100 km${\rm  s^{-1}}$. In fitting the grism spectra, we adopt an intrinsic 1D line profile.


\subsection{Confirmation of HAEs}
\label{sec:sample}

\begin{figure*}[t]
\centering  
\subfigure{
\includegraphics[width=0.44\linewidth]{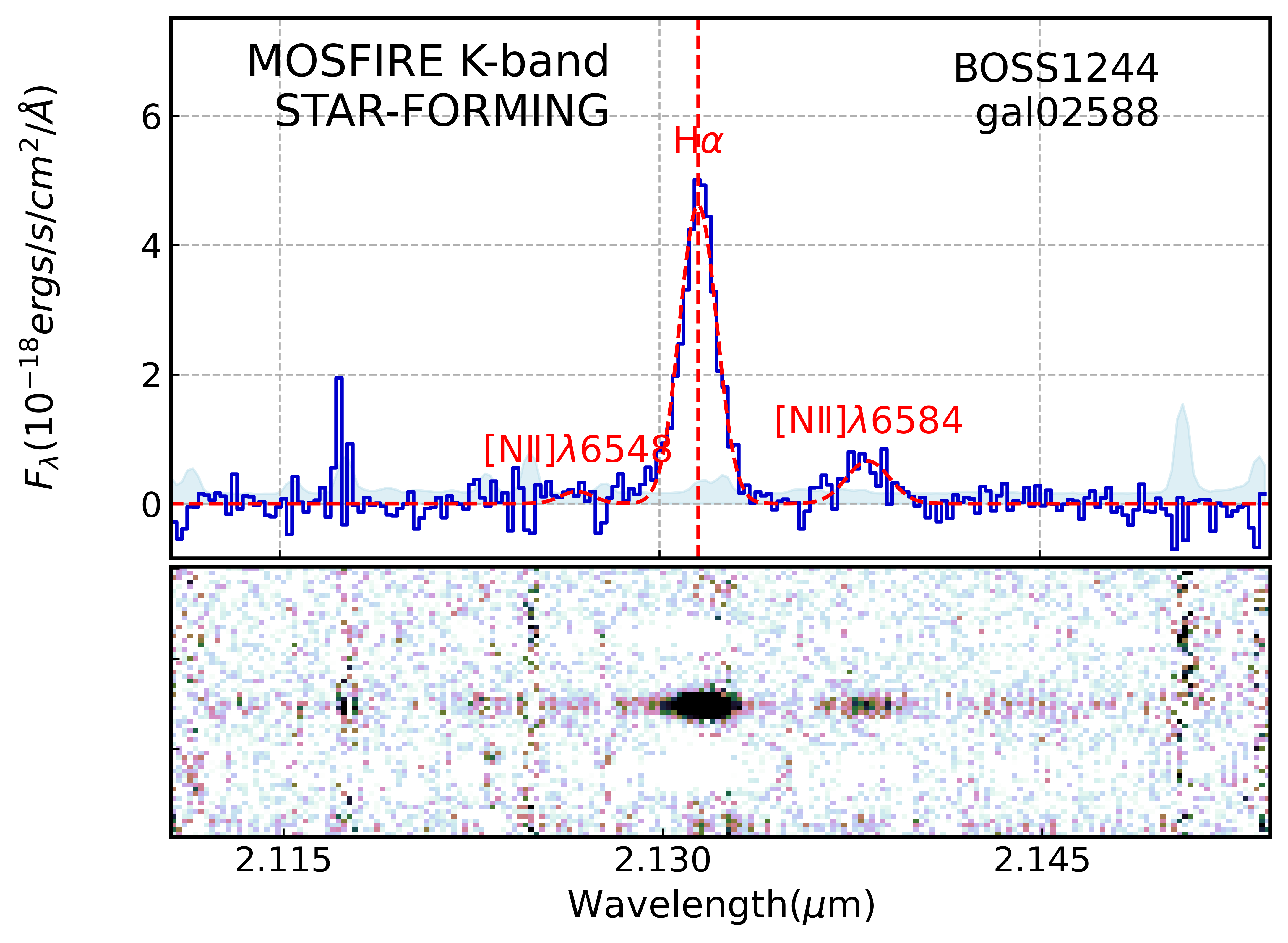}}
\subfigure{
\includegraphics[width=0.44\linewidth]{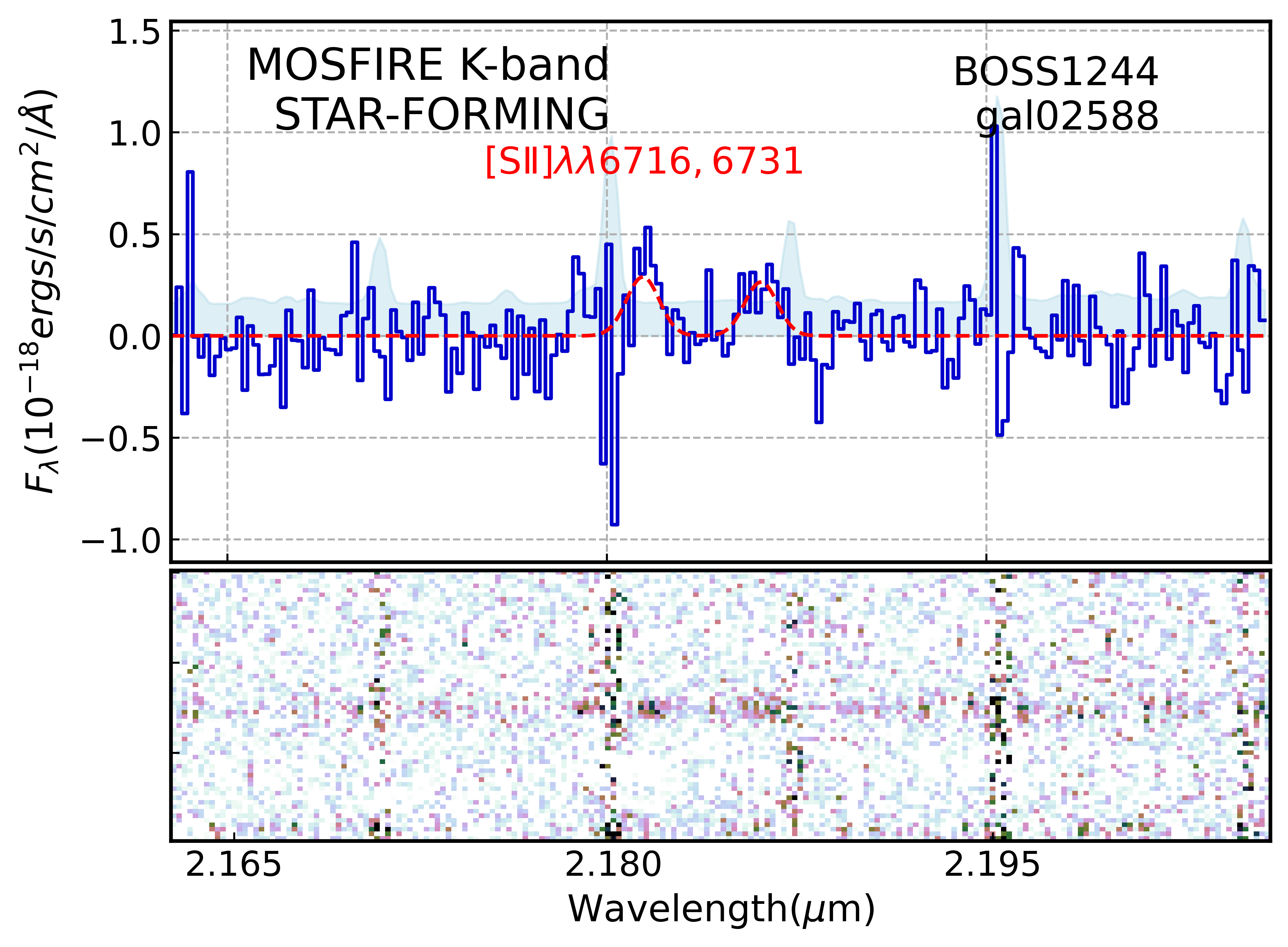}}
\caption{Flux-calibrated K-band spectrum for object gal02588, as listed in Table~\ref{table::measure}, observed with the Keck MOSFIRE instrument. The color-coding and styles of the symbols and lines follow exactly those shown in Figure~\ref{fig:show_galaxy_grizli}. The left panel displays the covered emission lines of \Ha and \NII, whereas the right panel shows \SII.}
\label{fig:show_galaxy}
\end{figure*}

We selected galaxies based on the following spectroscopic criteria: 

1. The galaxy's \Ha emission line flux has a SNR of at least 20.

2. The redshift fall within the range $z \sim 2.2-2.3$ to ensure that the primary emission lines, such as \Ha, \SII and \NII, are within the K-band, and the G141 grism can cover emission lines including \OIII, \Hb and \OII.


Figure~\ref{fig:show_galaxy} presents example spectra near the \Ha wavelength region for one confirmed HAE out of a total of 43. The left panel shows the 1D and 2D spectra of galaxy ID gal02588 in the BOSS1244 field covering emission line \Ha and \NII, while the right panel shows emission line \SII. The bottom panels display the 2D spectra, and the top panels show the corresponding 1D spectra, with the continuum subtracted. 
The blue shaded region indicates the error range of the observed spectrum, and the red lines represent the emission line models fitted using Gaussian function, as described in Section~\ref{sec:fitting}.
The 2D spectra clearly show \Ha emission lines, which serve as one of the criteria for identifying HAEs. 
Galaxy ID gal02588 exhibits strong \Ha emission along with clearly detected \NII lines in both the 2D and 1D spectra.

\subsection{Spectral Stacking}\label{sec::stack}

To investigate differences in the ionization mechanisms of ISM across galaxies with varying stellar masses, we analyzed both individual galaxy spectra and stacked spectra in different stellar mass bins. Stacked spectra enhances the SNR, enabling more precise measurements of weak emission line fluxes. For example, the \OI emission line is crucial to study ISM ionization mechanisms, as it can probe the nature of low-ionization regions and processes such as shocks. However, since it is a weak line, the \OI signal in individual galaxy spectra is often faint. By stacking spectra within different stellar mass intervals, the improved SNR allows for more robust measurements of average \OI flux. We divide these galaxies into two mass bins: $\rm log(M_{*}/M_{\odot}) \in$ [9.0,10.0), and $\rm log(M_{*}/M_{\odot}) \in$ [10.0,11.0]. Based on the stellar masses listed in Table~\ref{table::measure}, 12 galaxies fall into the low-mass bin, while 31 galaxies are classified into the high-mass bin.

We then adopt the following stacking procedures, following the method described in \cite{Wang_2022}, 
\begin{enumerate}
    \item Subtract continuum from the observed spectra. The K-band continuum spectrum is modeled by fitting a polynomial using the SPECUTILS \citep{Specutils,substract} software, while the G141-band continuum spectrum is constructed with the GRIZLI software. 

    \item Normalize the spectrum of each galaxy by measured \Ha emission line flux.

    \item Mask bright sky lines from the observed spectra to minimize contamination.
    
    \item Apply redshift correction to recover the rest-frame spectral wavelengths.

    \item Re-create the stacked spectra through non-parametric bootstrap resampling, drawing 1000 samples to represent the overall dataset. We take the median spectrum as the final stacked spectrum and  and the 1$\sigma$ uncertainty range is defined by the 16th and 84th percentiles.
\end{enumerate}
We additionally stack the spectra of all 43 galaxies.

\begin{figure*}[h]
\centering
\begin{minipage}{0.95\linewidth}
    \includegraphics[width=\linewidth]{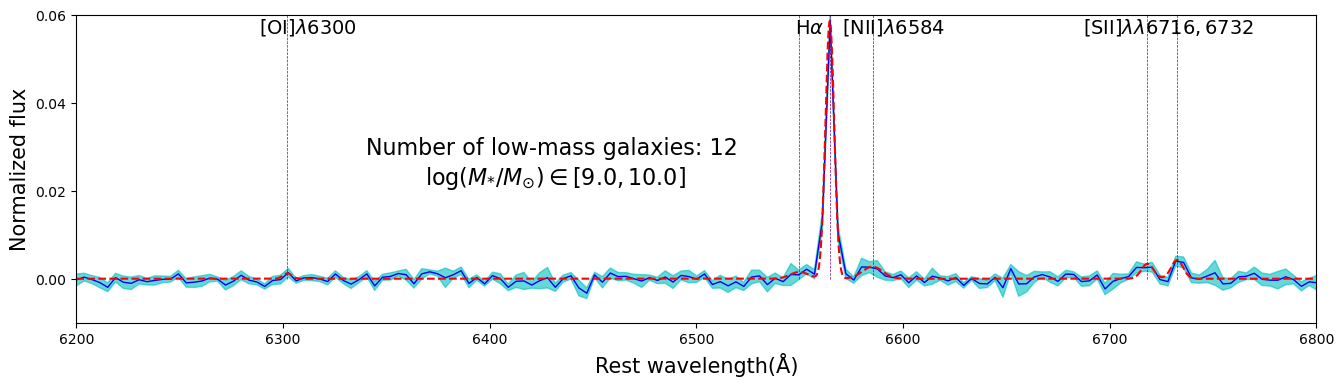}
\end{minipage}
\begin{minipage}{0.95\linewidth}
    \includegraphics[width=\linewidth]{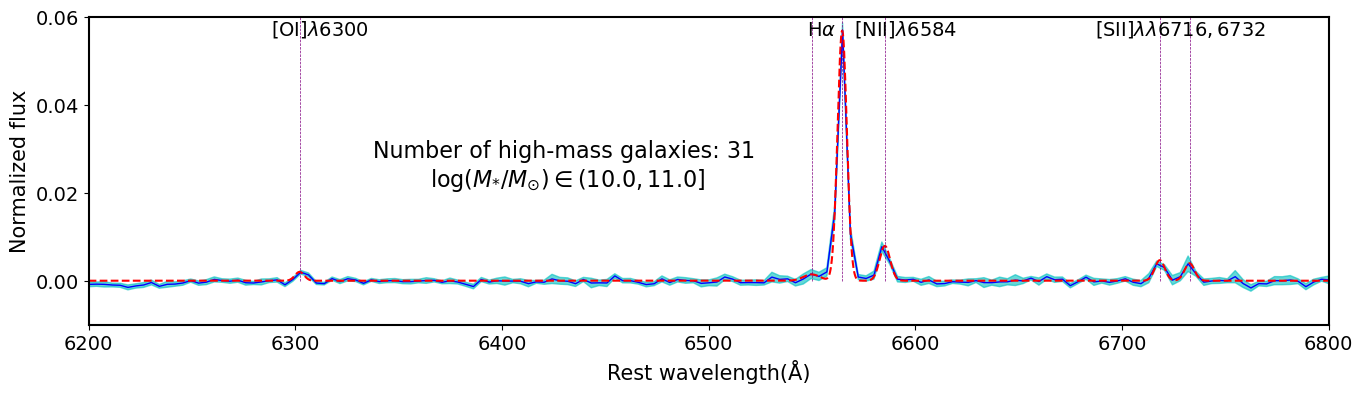}
\end{minipage}
\begin{minipage}{0.95\linewidth}
    \includegraphics[width=\linewidth]{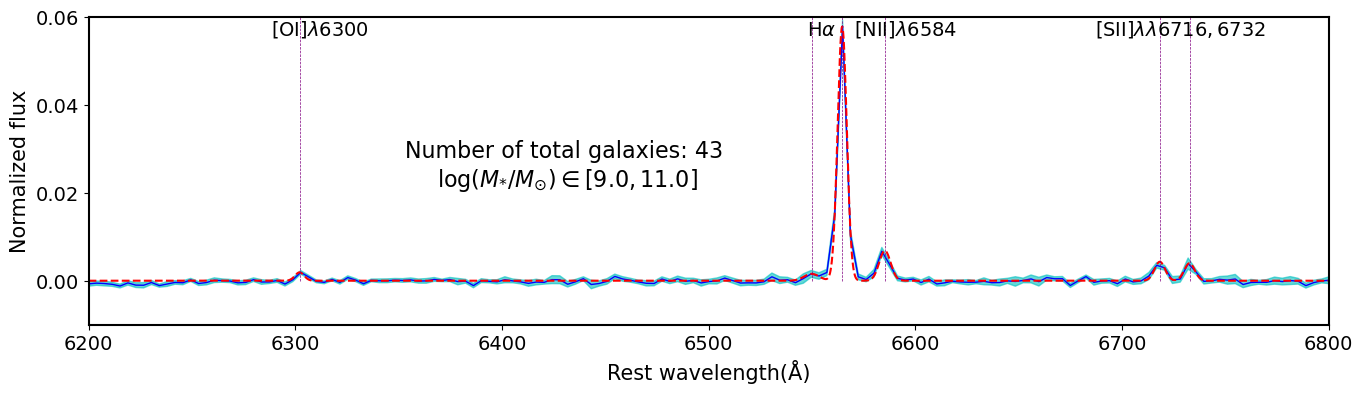}
\end{minipage}
\caption{Stacked 1D MOSFIRE spectra for 43 galaxies at $z\sim$2.2 in three massive protocluster fields (BOSS1244, BOSS1441, and BOSS1542). In the spectral plot, the blue curve shows the bootstrapped spectrum, the cyan bands indicate the bootstrapped flux uncertainties, and the red dashed curves show the best-fit Gaussian fits to multiple emission lines (\Ha, \NII, \OI, \SII). The details of the stacking procdeures are provided in the Section~\ref{sec::stack}.}
\label{fig:stacking_keck}
\end{figure*}

 We fit the emission lines in the three stacked spectra to derive the line fluxes, following the method described in Section~\ref{sec:fitting}. 
The stacked spectra and the emission lines models are shown in Figure~\ref{fig:stacking_keck}. From top to bottom, the panels show the stacked spectrum for the low-mass bin, the high-mass bin, and the 43 full stacked galaxies. The blue lines represent the spectra, while the blue shaded regions indicate the 1$\sigma$ error ranges. The red dashed lines correspond to the emission line models.

Notably, the \Ha lines are prominent in all three spectra, whereas the \OI, \NII, and \SII lines are clearly visible in the high-mass bin and the total stacked spectra.


\begin{turnpage}
\begin{deluxetable}{lccccccccccccccccc}
    \tablecolumns{11}
    \tablewidth{0pt}
    \label{table::measure}
    \tablecaption{Measured Properties of Individual Sources in Galaxies Residing in Extreme Overdensities at $z\sim2.2$.}
    \tablehead{
\colhead{Protocluster} & 
\colhead{\multirow{2}{*}{ID}}  & 
\colhead{\multirow{2}{*}{$z_{\rm spec}$}}   & 
\colhead{\multirow{2}{*}{$\rm\log\frac{[O\,{ III}]}{H\beta}$}} &
\colhead{\multirow{2}{*}{$\rm\log\frac{[N\,{ II}]}{H\alpha}$}} & 
\colhead{\multirow{2}{*}{$\rm\log\frac{[S\,{ II}]}{H\alpha}$}} & 
\colhead{\multirow{2}{*}{$\rm\log\frac{[O\,{ I}]}{H\alpha}$}} & 
\colhead{log($n_e $)} &
\colhead{stellar mass} & 
\colhead{12+log(O/H)}  &
\colhead{SFR} & 
\colhead{\SII}  & 
\colhead{$\frac{\OIII\lambda\lambda4959.5007)}{\OII\lambda\lambda3726.2729} $}\\
Field &  & &  & & &  &cm$^{-3}$&$\log(M_*/M_\odot)$ & & $\rm[M_*/yr]$ &  $\lambda6716/\lambda6731$ \\
}
\startdata
BOSS1244	&	gal00764	&2.22&	$	0.14_{-0.16}^{+0.12}	$	&	$	-0.78_{-0.07}^{+0.06}	$	&	$	-0.84_{-0.04}^{+0.04}	$	&	$	-	$	&$2.83_{-0.3}^{+0.3}$&10.52&$8.6_{-0.04}^{+0.05}$&$123.09_{-24.88}^{+24.88}$&$0.92_{-0.75}^{+1.13}$&$0.81_{-0.19}^{+0.29}$\\
BOSS1244	&	gal01335	&2.24&	$	0.41_{-0.09}^{+0.07}	$	&	$	-0.76_{-0.03}^{+0.03}	$	&	$	-0.45_{-0.03}^{+0.03}	$	&	$	-	$	&$2.82_{-0.23}^{+0.23}$&10.28&$8.59_{-0.02}^{+0.02}$&$5.25_{-0.35}^{+0.35}$&$0.92_{-0.8}^{+1.07}$&$0.82_{-0.09}^{+0.1}$\\
BOSS1244	&	gal01394	&2.22&	$	0.42_{-0.04}^{+0.03}	$	&	$	-0.84_{-0.11}^{+0.09}	$	&	$	-1.04_{-0.05}^{+0.04}	$	&	$	-1.29_{-0.08}^{+0.07}	$	&$2.79_{-0.34}^{+0.34}$&9.65&$8.41_{-0.01}^{+0.01}$&$108.44_{-8.60}^{+8.60}$&$0.93_{-0.74}^{+1.15}$&$1.66_{-0.12}^{+0.15}$\\
BOSS1244	&	gal01998	&2.21&	$	0.61_{-0.09}^{+0.07}	$	&	$	-0.94_{-0.05}^{+0.04}	$	&	$	-0.61_{-0.03}^{+0.02}	$	&	$	-1.29_{-0.07}^{+0.06}	$	&$3.19_{-0.15}^{+0.18}$&10.45&$8.3_{-0.02}^{+0.02}$&$13.37_{-1.18}^{+1.18}$&$0.72_{-0.64}^{+0.8}$&$2.54_{-0.33}^{+0.42}$\\
BOSS1244	&	gal02474	&2.19&	$	0.67_{-0.54}^{+0.23}	$	&	$	-0.59_{-0.04}^{+0.04}	$	&	$	-0.92_{-0.09}^{+0.08}	$	&	$	-	$	&$-$&9.93&$8.41_{-0.03}^{+0.03}$&$56.68_{-3.86}^{+3.86}$&$0.02_{-0.02}^{+0.19}$&$1.66_{-0.32}^{+0.43}$\\
BOSS1244	&	gal02570	&2.25&	$	0.07_{-0.23}^{+0.15}	$	&	$	-0.44_{-0.01}^{+0.01}	$	&	$	-0.63_{-0.03}^{+0.03}	$	&	$	-1.96_{-0.22}^{+0.14}	$	&$-$&10.4&$8.6_{-0.06}^{+0.07}$&$213.93_{-28.01}^{+28.01}$&$1.46_{-1.25}^{+1.69}$&$0.78_{-0.27}^{+0.32}$\\
BOSS1244	&	gal02611	&2.23&	$	0.44_{-0.16}^{+0.12}	$	&	$	-0.82_{-0.06}^{+0.05}	$	&	$	-0.86_{-0.06}^{+0.06}	$	&	$	-	$	&$3.3_{-0.4}^{+0.88}$&9.72&$8.31_{-0.05}^{+0.05}$&$51.07_{-3.21}^{+3.21}$&$0.66_{-0.45}^{+0.89}$&$2.51_{-0.69}^{+1.38}$\\
BOSS1244	&	gal02771	&2.24&	$	-	$	&	$	-	$	&	$	-0.43_{-0.04}^{+0.03}	$	&	$	-0.89_{-0.23}^{+0.15}	$	&$-$&9.79&$8.27_{-0.03}^{+0.04}$&$51.98_{-5.81}^{+5.81}$&$0.35_{-0.26}^{+0.43}$&$2.84_{-0.61}^{+1.0}$\\
BOSS1244	&	gal90044	&2.24&	$	-	$	&	$	-0.71_{-0.03}^{+0.02}	$	&	$	-0.69_{-0.02}^{+0.02}	$	&	$	-1.35_{-0.07}^{+0.06}	$	&$1.41_{-1.41}^{+0.78}$&10.1&$8.27_{-0.03}^{+0.04}$&$130.57_{-3.39}^{+3.39}$&$1.39_{-1.24}^{+1.57}$&$2.9_{-0.64}^{+1.06}$\\
BOSS1244	&	gal01467	&2.21&	$	0.7_{-0.04}^{+0.04}	$	&	$	-1.2_{-0.06}^{+0.05}	$	&	$	-0.71_{-0.03}^{+0.03}	$	&	$	-1.52_{-0.1}^{+0.08}	$	&$3.13_{-0.19}^{+0.23}$&10.15&$8.15_{-0.02}^{+0.02}$&$6.49_{-0.34}^{+0.34}$&$0.74_{-0.63}^{+0.86}$&$4.49_{-0.47}^{+0.59}$\\
BOSS1244	&	gal01855	&2.25&	$	0.26_{-0.09}^{+0.08}	$	&	$	-0.35_{-0.01}^{+0.01}	$	&	$	-0.68_{-0.03}^{+0.02}	$	&	$	-1.33_{-0.04}^{+0.03}	$	&$-$&10.76&$8.29_{-0.05}^{+0.06}$&$14.73_{-1.05}^{+1.05}$&$1.54_{-1.34}^{+1.81}$&$2.63_{-0.8}^{+2.16}$\\
BOSS1244	&	gal01890	&2.25&	$	0.92_{-0.44}^{+0.21}	$	&	$	<-1.84	$	&	$	-1.15_{-0.19}^{+0.13}	$	&	$	-	$	&$-$&9.99&$8.29_{-0.02}^{+0.03}$&$4.07_{-0.25}^{+0.25}$&$1.97_{-1.97}^{+5.69}$&$2.64_{-0.38}^{+0.59}$\\
BOSS1244	&	gal02992	&2.35&	$	0.45_{-0.06}^{+0.05}	$	&	$	-0.57_{-0.02}^{+0.02}	$	&	$	-0.92_{-0.06}^{+0.05}	$	&	$	-0.97_{-0.03}^{+0.03}	$	&$-$&10.31&$8.38_{-0.02}^{+0.02}$&$12.10_{-1.10}^{+1.10}$&$0.45_{-0.28}^{+0.61}$&$1.87_{-0.2}^{+0.22}$\\
BOSS1244	&	gal03061	&2.23&	$	0.56_{-0.05}^{+0.04}	$	&	$	-1.34_{-0.15}^{+0.11}	$	&	$	-1.12_{-0.05}^{+0.05}	$	&	$	-1.23_{-0.07}^{+0.06}	$	&$-$&10.21&$8.37_{-0.01}^{+0.01}$&$13.55_{-0.59}^{+0.59}$&$3.26_{-2.25}^{+5.78}$&$1.96_{-0.14}^{+0.17}$\\
BOSS1244	&	gal03331	&2.23&	$	0.95_{-0.57}^{+0.24}	$	&	$	<-1.74	$	&	$	-1.06_{-0.08}^{+0.07}	$	&	$	-1.0_{-0.17}^{+0.12}	$	&$-$&9.95&$8.31_{-0.02}^{+0.03}$&$9.17_{-0.54}^{+0.54}$&$4.37_{-4.37}^{+14.44}$&$2.42_{-0.37}^{+0.45}$\\
BOSS1244	&	gal00306	&2.22&	$	-	$	&	$	-0.44_{-0.01}^{+0.01}	$	&	$	-0.78_{-0.02}^{+0.02}	$	&	$	-1.38_{-0.23}^{+0.15}	$	&$1.91_{-1.91}^{+0.44}$&10.96&$-$&$27.70_{-2.56}^{+2.56}$&$1.32_{-1.18}^{+1.48}$&$-$\\
BOSS1244	&	gal00327	&2.23&	$	0.69_{-0.25}^{+0.16}	$	&	$	-1.4_{-0.27}^{+0.17}	$	&	$	-1.21_{-0.12}^{+0.09}	$	&	$	-0.94_{-0.04}^{+0.04}	$	&$-$&9.86&$8.47_{-0.02}^{+0.02}$&$6.11_{-0.26}^{+0.26}$&$-$&$1.3_{-0.15}^{+0.21}$\\
BOSS1244	&	gal00613	&2.23&	$	0.63_{-0.14}^{+0.1}	$	&	$	-1.5_{-0.12}^{+0.1}	$	&	$	-1.09_{-0.1}^{+0.08}	$	&	$	-1.53_{-0.13}^{+0.1}	$	&$-$&10.1&$8.34_{-0.03}^{+0.03}$&$9.57_{-0.41}^{+0.41}$&$3.35_{-3.35}^{+10.3}$&$2.23_{-0.33}^{+0.43}$\\
BOSS1244	&	gal01417	&2.25&	$	0.26_{-0.1}^{+0.08}	$	&	$	-0.8_{-0.03}^{+0.02}	$	&	$	-0.59_{-0.03}^{+0.03}	$	&	$	-1.34_{-0.45}^{+0.22}	$	&$-$&10.3&$-$&$13.33_{-0.71}^{+0.71}$&$1.62_{-1.39}^{+1.88}$&$3.51_{-14.6}^{+7.73}$\\
BOSS1244	&	gal01878	&2.06&	$	0.25_{-0.06}^{+0.05}	$	&	$	-1.05_{-0.06}^{+0.05}	$	&	$	-0.57_{-0.04}^{+0.04}	$	&	$	-	$	&$2.1_{-2.1}^{+0.51}$&10.12&$8.51_{-0.02}^{+0.02}$&$6.77_{-0.48}^{+0.48}$&$1.25_{-1.02}^{+1.51}$&$1.13_{-0.14}^{+0.19}$\\
BOSS1244	&	gal02327	&2.24&	$	0.64_{-0.08}^{+0.07}	$	&	$	-1.54_{-0.18}^{+0.12}	$	&	$	-0.74_{-0.03}^{+0.03}	$	&	$	-	$	&$2.9_{-0.22}^{+0.24}$&10.21&$8.37_{-0.01}^{+0.02}$&$10.17_{-0.48}^{+0.48}$&$0.88_{-0.75}^{+1.02}$&$1.96_{-0.17}^{+0.22}$\\
BOSS1244	&	gal02588	&2.25&	$	0.38_{-0.11}^{+0.09}	$	&	$	-0.85_{-0.02}^{+0.02}	$	&	$	-0.86_{-0.04}^{+0.04}	$	&	$	-1.55_{-0.08}^{+0.07}	$	&$2.69_{-0.32}^{+0.29}$&10.36&$8.52_{-0.02}^{+0.02}$&$18.44_{-1.12}^{+1.12}$&$0.98_{-0.82}^{+1.18}$&$1.08_{-0.11}^{+0.14}$\\
BOSS1244	&	gal02950	&2.23&	$	0.45_{-0.15}^{+0.11}	$	&	$	-1.14_{-0.33}^{+0.18}	$	&	$	-0.99_{-0.07}^{+0.06}	$	&	$	<-1.23	$	&$-$&10.53&$8.36_{-0.03}^{+0.04}$&$6.98_{-1.54}^{+1.54}$&$-$&$2.02_{-0.4}^{+0.71}$\\
BOSS1244	&	gal03040	&2.25&	$	-	$	&	$	<-2.71	$	&	$	-0.96_{-0.06}^{+0.06}	$	&	$	-1.38_{-0.13}^{+0.1}	$	&$-$&10.16&$-$&$7.25_{-0.54}^{+0.54}$&$-$&$-$\\
BOSS1244	&	gal03340	&2.23&	$	0.58_{-0.27}^{+0.17}	$	&	$	-0.71_{-0.03}^{+0.03}	$	&	$	-0.66_{-0.04}^{+0.04}	$	&	$	-1.1_{-0.06}^{+0.05}	$	&$2.52_{-0.48}^{+0.35}$&10.18&$8.39_{-0.04}^{+0.04}$&$5.67_{-0.26}^{+0.26}$&$1.09_{-0.9}^{+1.33}$&$1.79_{-0.41}^{+0.68}$\\
BOSS1244	&	gal01070	&2.23&	$	0.93_{-0.48}^{+0.22}	$	&	$	-	$	&	$	-1.0_{-0.16}^{+0.12}	$	&	$	-1.27_{-0.22}^{+0.15}	$	&$-$&9.96&$8.26_{-0.03}^{+0.04}$&$40.86_{-1.74}^{+1.74}$&$0.06_{-0.06}^{+0.35}$&$2.92_{-0.65}^{+1.11}$\\
BOSS1244	&	gal02510	&2.23&	$	0.09_{-0.06}^{+0.05}	$	&	$	-0.69_{-0.01}^{+0.01}	$	&	$	-0.69_{-0.02}^{+0.02}	$	&	$	-1.36_{-0.05}^{+0.04}	$	&$2.94_{-0.14}^{+0.15}$&10.34&$8.43_{-0.02}^{+0.02}$&$16.79_{-0.29}^{+0.29}$&$0.85_{-0.77}^{+0.94}$&$1.53_{-0.2}^{+0.24}$\\
BOSS1441	&	gal03745	&2.25&	$	0.69_{-0.11}^{+0.08}	$	&	$	-1.04_{-0.08}^{+0.07}	$	&	$	-1.09_{-0.52}^{+0.23}	$	&	$	-	$	&$-$&10.4&$8.22_{-0.02}^{+0.03}$&$111.86_{-12.37}^{+12.37}$&$0.08_{-0.08}^{+0.92}$&$3.47_{-0.54}^{+0.79}$\\
BOSS1441	&	gal04511	&2.21&	$	0.13_{-0.07}^{+0.06}	$	&	$	-0.78_{-0.14}^{+0.11}	$	&	$	-0.63_{-0.09}^{+0.08}	$	&	$	-1.81_{-1.87}^{+0.3}	$	&$3.35_{-0.57}^{+0.8}$&10.1&$8.56_{-0.02}^{+0.03}$&$105.75_{-9.17}^{+9.17}$&$0.65_{-0.31}^{+0.99}$&$0.91_{-0.13}^{+0.15}$\\
BOSS1441	&	gal04563	&2.31&	$	<1.2	$	&	$	-1.68_{-0.18}^{+0.13}	$	&	$	-1.34_{-0.14}^{+0.11}	$	&	$	-1.97_{-0.38}^{+0.2}	$	&$-$&9.89&$8.25_{-0.02}^{+0.02}$&$85.06_{-5.32}^{+5.32}$&$2.19_{-2.19}^{+7.43}$&$3.11_{-0.38}^{+0.5}$\\
\enddata

\end{deluxetable}

\end{turnpage}

\begin{turnpage}

\begin{deluxetable*}{lccccccccccccccccc}
    \tablecolumns{11}
    \tablewidth{0pt}
    \label{table:measure}
    \setcounter{table}{2}
    \tablecaption{(\textit{continued})}
    \tablehead{
\colhead{Protocluster} & 
\colhead{\multirow{2}{*}{ID}}  & 
\colhead{\multirow{2}{*}{$z_{\rm spec}$}}   & 
\colhead{\multirow{2}{*}{$\rm\log\frac{[O\,{ III}]}{H\beta}$}} &
\colhead{\multirow{2}{*}{$\rm\log\frac{[N\,{ II}]}{H\alpha}$}} & 
\colhead{\multirow{2}{*}{$\rm\log\frac{[S\,{ II}]}{H\alpha}$}} & 
\colhead{\multirow{2}{*}{$\rm\log\frac{[O\,{ I}]}{H\alpha}$}} & 
\colhead{log($n_e $)} &
\colhead{stellar mass} & 
\colhead{12+log(O/H)}  &
\colhead{SFR} & 
\colhead{\SII}  & 
\colhead{$\frac{\OIII\lambda\lambda4959.5007)}{\OII\lambda\lambda3726.2729} $}\\
Field &  & &  & & &  &cm$^{-3}$&$\log(M_*/M_\odot)$ & & $\rm[M_*/yr]$ &  $\lambda6716/\lambda6731$ \\
}
\startdata
BOSS1441	&	gal04794	&2.34&	$	0.59_{-0.34}^{+0.19}	$	&	$	-1.23_{-0.15}^{+0.11}	$	&	$	-1.12_{-0.12}^{+0.09}	$	&	$	-1.17_{-0.28}^{+0.17}	$	&$-$&10.64&$8.49_{-0.03}^{+0.03}$&$66.94_{-9.41}^{+9.41}$&$-$&$1.21_{-0.21}^{+0.28}$\\
BOSS1441	&	gal04932	&2.31&	$	0.47_{-0.15}^{+0.10}	$	&	$	-0.6_{-0.11}^{+0.09}	$	&	$	-0.69_{-0.07}^{+0.06}	$	&	$	-	$	&$-$&10.2&$8.28_{-0.01}^{+0.01}$&$104.14_{-3.83}^{+3.83}$&$3.53_{-2.65}^{+4.81}$&$2.77_{-0.21}^{+0.29}$\\
BOSS1542	&	gal04669	&2.27&	$	-	$	&	$	-0.88_{-0.1}^{+0.08}	$	&	$	-0.41_{-0.04}^{+0.04}	$	&	$	-1.43_{-0.33}^{+0.18}	$	&$2.6_{-0.42}^{+0.34}$&10.2&$-$&$3.18_{-0.13}^{+0.13}$&$1.02_{-0.83}^{+1.26}$&$-$\\
BOSS1542	&	gal04816	&2.35&	$	-	$	&	$	-0.73_{-0.02}^{+0.02}	$	&	$	-0.97_{-0.05}^{+0.05}	$	&	$	-1.59_{-0.12}^{+0.09}	$	&$-$&9.12&$-$&$8.61_{-0.18}^{+0.18}$&$3.85_{-2.67}^{+7.02}$&$-$\\
BOSS1542	&	gal05204	&2.20&	$	-0.06_{-0.26}^{+0.16}	$	&	$	-0.41_{-0.05}^{+0.04}	$	&	$	-0.75_{-0.12}^{+0.09}	$	&	$	-	$	&$-$&9.52&$8.56_{-0.06}^{+0.07}$&$4.47_{-0.22}^{+0.22}$&$0.24_{-0.24}^{+0.51}$&$0.91_{-0.34}^{+0.9}$\\
BOSS1542	&	gal90082	&2.23&	$	-	$	&	$	-1.0_{-0.03}^{+0.03}	$	&	$	-0.94_{-0.05}^{+0.04}	$	&	$	-1.18_{-0.05}^{+0.05}	$	&$1.58_{-1.58}^{+1.0}$&10.1&$8.57_{-0.06}^{+0.09}$&$6.08_{-0.20}^{+0.20}$&$1.33_{-1.05}^{+1.76}$&$0.88_{-0.36}^{+0.97}$\\
BOSS1542	&	gal01338	&2.25&	$	0.37_{-0.12}^{+0.09}	$	&	$	-0.76_{-0.04}^{+0.04}	$	&	$	-0.79_{-0.15}^{+0.11}	$	&	$	-1.3_{-0.11}^{+0.09}	$	&$-$&10.2&$8.45_{-0.04}^{+0.04}$&$4.76_{-0.84}^{+0.84}$&$1.78_{-1.78}^{+4.71}$&$1.42_{-0.31}^{+0.51}$\\
BOSS1542	&	gal02830	&2.23&	$	-	$	&	$	-1.26_{-0.1}^{+0.08}	$	&	$	-0.68_{-0.04}^{+0.04}	$	&	$	-1.65_{-0.26}^{+0.16}	$	&$2.65_{-0.36}^{+0.31}$&9.61&$-$&$4.99_{-1.11}^{+1.11}$&$1.01_{-0.84}^{+1.23}$&$-$\\
BOSS1542	&	gal02872	&2.24&	$	-	$	&	$	-0.68_{-0.04}^{+0.03}	$	&	$	-0.75_{-0.06}^{+0.05}	$	&	$	-	$	&$-$&10.24&$-$&$5.93_{-0.80}^{+0.80}$&$3.47_{-2.48}^{+5.23}$&$-$\\
BOSS1542	&	gal04010	&2.25&	$	-	$	&	$	-0.52_{-0.04}^{+0.04}	$	&	$	-0.13_{-0.05}^{+0.04}	$	&	$	-	$	&$3.38_{-0.24}^{+0.37}$&10.59&$8.39_{-0.03}^{+0.04}$&$7.37_{-0.70}^{+0.70}$&$0.63_{-0.53}^{+0.77}$&$1.78_{-0.41}^{+0.71}$\\
BOSS1542	&	gal01145	&2.25&	$	-	$	&	$	-	$	&	$	-0.11_{-0.1}^{+0.08}	$	&	$	-	$	&$-$&10.06&$-$&$3.41_{-0.30}^{+0.30}$&$0.02_{-0.02}^{+0.05}$&$-$\\
BOSS1542	&	gal01704	&2.34&	$	-	$	&	$	-0.72_{-0.04}^{+0.04}	$	&	$	-0.74_{-0.07}^{+0.06}	$	&	$	-0.99_{-0.06}^{+0.05}	$	&$2.94_{-0.53}^{+0.68}$&10.06&$8.18_{-0.04}^{+0.05}$&$8.96_{-0.66}^{+0.66}$&$0.86_{-0.58}^{+1.2}$&$4.14_{-0.98}^{+1.53}$\\
BOSS1542	&	gal03125	&2.31&	$	0.3_{-0.1}^{+0.08}	$	&	$	-0.75_{-0.03}^{+0.02}	$	&	$	-0.68_{-0.03}^{+0.03}	$	&	$	-1.39_{-0.28}^{+0.17}	$	&$-$&10.1&$8.18_{-0.04}^{+0.04}$&$10.01_{-1.03}^{+1.03}$&$1.71_{-1.46}^{+2.03}$&$4.0_{-0.85}^{+1.55}$\\
\enddata
\tablenotetext{a}{A dash ($-$) indicates that the line is too weak to be detected, and thus no corresponding value is available.}
\tablenotemark{b}{The metaillicity data are are based on the \citep{Bian_2018} calibrations.}
\end{deluxetable*}
\end{turnpage}
\clearpage

\section{Emission Line Diagnostics}
\label{sec:measure}

Our HST slitless spectra measure the \OII, \Hb and \OIII emission lines, while our Keck/MOSFIRE observations provide \Ha, \NII, \SII and \OI lines. These emission lines can be used to infer the average $n_e$, oxygen abundance, and ionization parameter $q$ of the ISM.

We calculated the SFRs using the method presented in \citet{Wang_2022}, which are listed in Table~\ref{table::measure}.
Following previous studies \citep{wangGrismLensAmplifiedSurvey2017,wangDiscoveryStronglyInverted2019,wangCensusSubkiloparsecResolution2020}, we jointly constrain gas-phase metallicity ($\rm12 + log(O/H)$), nebular dust extinction ($\rm A_{v}$), and dereddened \Hb flux using forward-modeling Bayesian inference method. Dust attenuation is corrected using the \citep{Calzetti_2000} extinction law, with $\rm A_{v}$ drawn from parameter sampling. SFRs are calculated using:
\begin{equation}\label{eq:wang2022}
\rm{SFR} =4.6\times 10^{-42}\frac{L(H{\alpha})}{erg\ s^{-1}}[M_{\odot}\rm{yr^{-1}}]
\end{equation}

\begin{deluxetable*}{lccccccccccccccccc}
    \tablecolumns{11}
    \tablewidth{0pt}
    \tablecaption{Measured Properties of the Stacked Spectra.}\label{table:stacking}
    \tablehead{
mass bin & N$_{gal}$ & log(M$_{*}$/M$_\odot$)  & $\rm\log\frac{\OIII}{H\beta}$ & $\rm\log\frac{\NII}{H\alpha}$ & $\rm\log\frac{\SII}{H\alpha}$ & $\rm\log\frac{\OI}{H\alpha}$ & log($n_e$) & 12+log(O/H)&  \SII  & $\frac{\OIII\lambda\lambda4959,5007}{\OII\lambda\lambda3726,2729}$  \\
 &  &   &  & &  &  & $\rm cm^{-3}$ &  &  $\lambda6716/\lambda6731$ & \\  \\
}

\startdata
low mass	&12&	[9.0-10.0]	&	$	0.76_{-0.11}^{+0.09}	$	&	$	-0.97_{-0.18}^{+0.13}	$	&	$	-0.76_{-0.06}^{+0.05}	$	&	$	-1.72_{-0.12}^{+0.09}	$	&$3.03_{-0.4}^{+0.25}$&	$	8.27_{-0.07}^{+0.24}	$	&	$0.80_{-0.19}^{+0.19}$	& 	$2.86_{-1.79}^{+1.09}$ \\
high mass	&31&	[10.0-11.0]	&	$	0.52_{-0.06}^{+0.06}	$	&	$	-0.75_{-0.03}^{+0.03}	$	&	$	-0.77_{-0.05}^{+0.04}	$	&	$	-1.44_{-0.05}^{+0.04}	$	&$1.97_{-1.97}^{+0.5}$&	$	8.29_{-0.07}^{+0.21}	$	&	$1.37_{-0.31}^{+0.31}$	&	$2.63_{-1.56}^{+0.85}$ \\
full stack	&43&	-	&	$	0.62_{-0.08}^{+0.07}	$	&	$	-0.82_{-0.04}^{+0.04}	$	&	$	-0.77_{-0.04}^{+0.03}	$	&	$	-1.48_{-0.04}^{+0.04}	$	&$2.47_{-0.75}^{+0.06}$&	$	8.28_{-0.07}^{+0.07}	$	& $1.15_{-0.20}^{+0.20}$	 & $2.75_{-0.57}^{+1.00}$	\\
\enddata
\tablenotetext{a}{The metaillicity data are based on the \citep{Bian_2018} calibrations.}
\end{deluxetable*}

The extinction curve is applied using \citep{Cardelli1989}. Details will be provided in Yang et al. (2025, in prep).
However, in this work, we only utilized the dust extinction component.
From this, we inferred the O32 ratio, which was then used to derive the corresponding metallicity.

\subsection{Electron Density}

Electron density is a key parameter in determining the intensity of collisionally excited forbidden emission lines. 
The flux ratios of density-sensitive emission lines, such as the \OII and \SII doublets, can be used to estimate the electron density in both local galaxies and those at high redshifts \citep[e.g.,][]{Bennert2006,Sanders_2016}.
We note that our spectral resolution is sufficient to fully resolved the \SII doublet components, but not those of the more closely spaced \OII doublets.

\cite{Sanders_2016} investigated the electron densities of star-forming galaxies at $z\sim2.3$ in the MOSFIRE Deep Evolution Field survey.
They adopted effective collision strengths calculated for an electron temperature of 10,000 K, which represents the typical equilibrium temperature of \HII regions. Although assuming a fixed electron temperature of 10,000 K may lead to an overestimation of electron density in metal-rich galaxies and an underestimation in metal-poor galaxies, the uncertainty introduced by this assumption is smaller than the typical measurement uncertainties for high-z galaxies.
We follow the standard method described in \cite{Sanders_2016} to measure electron density:
\begin{equation}\label{eq:sanders2016}
n_e(R) = \frac{cR - ab}{c - R}
\end{equation}
where $R=\SII \lambda6717/\SII \lambda6731$. The constants are $a=0.4315$, $b=2.107$, and $c=627.1$.

We calculate the electron densities for all 43 spectra, and stacked spectra. However, since the \SII lines are relatively weak, we ultimately determine the electron densities for only a subset of the sample.

The results, including uncertainties, are presented in Table~\ref{table::measure}, with stacked results summarized separately in Table~\ref{table:stacking}. 
We find that the electron density in the full stacked spectrum is about 290 $\rm cm^{-3}$, high-mass stacked sample is 93 $\rm cm^{-3}$ and low-mass stacked sample is 1070 $\rm cm^{-3}$. Individual galaxies exhibit electron densities ranging from 37 $\rm cm^{-3}$ to 2410 $\rm cm^{-3}$.

The average electron density in our galaxies is consistent with previous studies. \cite{Sanders_2016} reported electron densities of 260 cm$^{-3}$ and 291 cm$^{-3}$, similar to the results of \cite{Shimakawa15} for the $z=2.5$ protocluster, and 10 times higher than in local galaxies. The low-mass stacked galaxies have high electron density and the high-mass stacked galaxies have low electron density. Additionally, \cite{Steidel_2014} estimated \SII  ratio to measure an electron density of 243 cm$^{-3}$ from their stacking analysis of 113 spectra at $z\sim2.3$ using Equation~\ref{eq:sanders2016}. 


\subsection{Oxygen Abundance and Ionization Parameter}

There are various methods to measure gas-phase oxygen abundance, including the ``direct" ($T_e$) method and the empirical metallicity calibrations derived from strong emission lines \citep[e.g.,][]{Ly2014, Steidel_2014,Newman_2014,Liu_2008}. However, these methods are usually calibrated for local galaxies and may not be directly applicable to high-redshift galaxies. For our $z\sim2$ galaxies, we adopt the ``direct" metallicity method using O32 \citep{Bian_2018}.
\cite{Bian_2018} utilized stacked spectra of star-forming galaxies at $z\sim2$, detecting the weak auroral line $\OIII\lambda4363$ to determine the direct electron temperature ($T_e$). Based on this approach, they established an empirical relation between the direct oxygen abundance and strong-line ratios, such as O32. This calibration, particularly relevant for high-redshift star-forming galaxies, is:
\begin{equation}
\label{eq:n2}
   12+\rm{log(O/H)}=8.54-0.59\times \rm{O32}
\end{equation} 
where $\rm O32=log(\OIII\lambda\lambda4959,5007/\OII\lambda\lambda3727,3729)$.

The metallicities of 43 galaxies and those derived from the stacked spectra, are given in Tables~\ref{table::measure} and \label{table:stacking}, respectively. For seven galaxies, the lack of emission lines required to measure the O32 ratio prevented a metallicity determination. The metallicities of the remaining galaxies are relatively similar, spanning a range of 8.15 to 8.60. The metallicities derived from the stacked spectra are also consistent across different mass bins, with the low-mass bin showing a metallicity of 8.27$_{-0.07}^{+0.24}$ and the high-mass bin yielding 8.29$_{-0.07}^{+0.21}$.

We show the mass-metallicity relation (MZR) of our sample in Figure.~\ref{fig:metailicity}.
We observe that our sample galaxies are located below the MOSDEF sample \citep{Sanders_2021}, suggestive of more efficient gas dilution in our protocluster environments \citep{Wang_2022,Li_22}. Additionally, we include the protocluster sample from Yang et al. (2025, in prep.), which also shows that protocluster galaxies exhibit lower metallicity than normal field galaxies. By fixing the intercept and fitting our sample against the protocluster field galaxies, we find a difference of 0.03 in the slope comparing to the samples from Yang et al. (2025, in prep.). This suggests a minor selection bias in the protocluster sample selection.

We find that 80\% of the galaxies in the sample have metallicities below 8.5. This preference for a metal-poor ISM aligns with the expectation that star-forming galaxies at $z\sim2$ have lower metallicities compared to their local counterparts.

The ionization parameter ($q$) is a fundamental physical quantity 
defined as the ratio of the ionizing photon flux per unit area ($S_{H_0}$) to the local hydrogen number density ($n$):
\begin{equation}\label{ioniztion}
q = \frac{S_{H_0}}{n}
\end{equation}
where $S_{\rm H_0}$ represents the ionizing photon flux, and $n$ is the local hydrogen number density. $q$ measures the density of ionizing photons per unit volume and can also be interpreted as the maximum velocity of an ionization front driven by the local radiation field. 

\begin{deluxetable}{lcccccc}
    \label{table:ionztion}
    \tablecolumns{7}
    \tablewidth{0pt}
    \tablecaption{Parameters Used to Calculate the Ionization Parameter ($q$).}

    \tablehead{
        Metallicity\tablenotemark{a}&  & 0.1 Z$_{\odot}$ & 0.2 Z$_{\odot}$ & 0.5 Z$_{\odot}$ \\
    }
    \startdata
    & $k_{0}$  &  7.46218  &  7.57817  & 7.73013 & \\
    & $k_{1}$ & 0.685835    &  0.739315  & 0.843125 & \\
    & $k_{2}$   & 0.0866086   &  0.0843640  & 0.118166  & \\
        \enddata
    \tablenotemark{a}{The metallicity value refers to the metallicity used in the photoionization model.}
\end{deluxetable}



Using photoionization models, $q$ can be derived from the ratio log($\OIII\lambda5007$/$\OII\lambda\lambda3727,3729$), which is highly sensitive to the ionization parameter, as it reflects the relative populations of ions in different ionization states \citep{Kewley_2002}. The relationship between \OIII/\OII\ and $q$ is parameterized through polynomial fitting, with coefficients that depend on metallicity. The metallicity is first estimated, and an iterative approach is used to simultaneously solve for $q$ and metallicity:
\begin{equation}
\label{eq:n2}
   q=10^{(k_0+k_1*\rm{R}+k_2*\rm{R}^{2})}
\end{equation} 
where $R$ is the flux ratio log(\OIII/\OII), and $k_0,\ k_1$, and $k_2$ are coefficients dependent on metallicity. We adopt this relation to calculate $q$.

Since 80\% of the galaxies in our sample have metallicities 12 + log(O/H) $<$ 8.5, we compute the ionization parameter assuming three metallicities: 
0.1 Z$_\odot$, 0.2 Z$_\odot$, and 0.5 Z$_\odot$. Our analysis reveals the following:

\begin{figure}[htbp]
\centering
\includegraphics[width=1.0\linewidth]{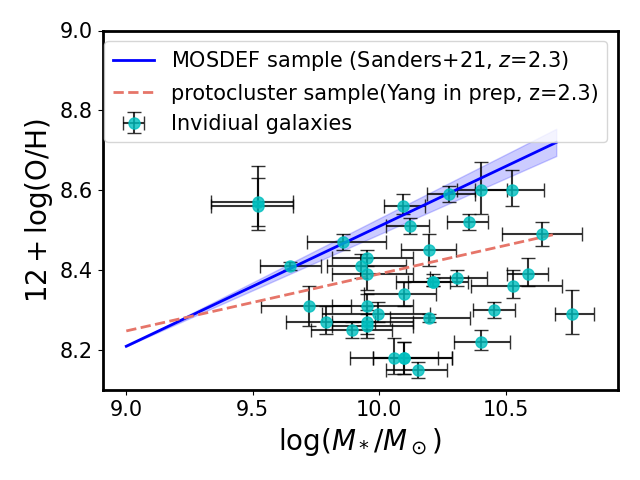}
\caption{The Mass-Metallicity Relation (MZR) for our galaxy sample is shown, where the cyan points represent individual high-redshift galaxies, with error bars indicating 1$\sigma$ uncertainties in the measurements. For comparison, the MOSDEF sample \citep{Sanders_2021} is plotted in blue, with light blue error bars denoting the uncertainties. Additionally, we include the protocluster sample from Yang et al. (2025, in prep.), marked by red dashed lines.}
\label{fig:metailicity}
\end{figure}

\begin{itemize}

    \item 
    At 0.1 Z$_\odot$, 80\% of the galaxies have log($q)>7.5$.
    
    \item 
    At 0.2 Z$_\odot$, 100\% of the galaxies exhibit log($q)>7.5$.
    
    \item 
    At 0.5 Z$_\odot$, 100\% of the galaxies have log($q)>7.5$.

\end{itemize}

For the stacked spectra, we applied the same metallicity assumptions and calculated the ionization parameters. We find consistently high log($q$) values, log($q)>7.75$ for all cases. These findings indicate that protoclusters exhibit elevated ionization parameters. 
This enhanced ionization parameter is similar to what is found in field galaxy spectra at $z$ = 1.5 \citep{Hayashi2015, Battisti2024}.

\subsection{O3R2 Diagram}
\label{sec:O32}

\begin{figure}[htbp]
\centering
\includegraphics[width=1.0\linewidth]{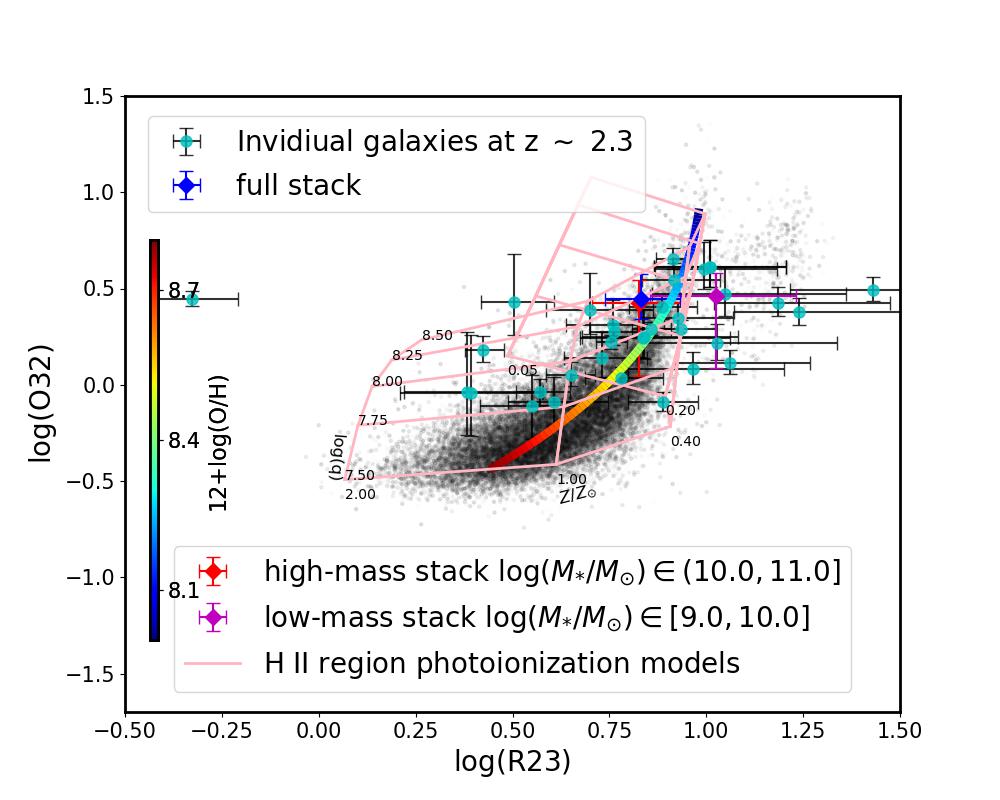}
\caption{O32 vs. R23 diagram for our sample galaxies. The cyan points indicate individual high-redshift galaxies from the sample, while the blue point represents the stacked sample of all galaxies. The purple point corresponds to the stacked sample of low-mass galaxies, and the red point denotes the stacked sample of high-mass galaxies. Error bars represent uncertainties in the measurements. The grey background shows the line ratios of local galaxies from the Sloan Digital Sky Survey (SDSS) for comparison. The colorbar adopts the calibration described in \cite{Bian_2018}. The pure \HII region photoionization model, featuring varying ionization parameters and metallicities, is shown in pink.
}
\label{fig:O32}
\end{figure}

In addition to electron density, metallicity, and ionization parameter, the BPT diagrams and the O32-R23 excitation diagram (O3R2 diagram) are critical related tools for examining the ionization properties of a galaxy's ISM. The O32 ratio ($\frac{\OIII\lambda\lambda4959,5007}{\OII\lambda\lambda3727,3729}$) and R23 ratio ($\frac{\OIII\lambda\lambda4959,5007+\OII\lambda\lambda3727,3729}{\rm{H}\beta}$) are highly sensitive to both metallicity and ionization parameter
\citep{Kewley_2002,Shapley15}.

High-redshift galaxies ($z\sim2.2$) have higher O32 and R23 values compared to local galaxies, occupying the tail region of the local distribution characterized by low metallicity and high ionization parameters \citep{Shapley15}. 
The high ionization parameter is defined based on our model grid and is illustrated in Figure.~\ref{fig:O32}, where less than half of the SDSS sample lies above this value. Accordingly, we consider ionization parameters above 7.5 to be high. Since our grid spans a ionization parameter range from 6.5 to 9.0, we adopt 7.5 as the threshold for a high ionization parameter.

In this work, we use the O3R2 diagram to examine the metallicities and ionization parameters of the sample galaxies, as shown in Figure~\ref{fig:O32}. The grey points represent local galaxies from the Sloan Digital Sky Survey (SDSS) Data Release 7 (DR7) catalog \citep{sdssdr7}, while the cyanpoints denote the 43 HAEs identified in this work. The red, purple, and blue rhombic symbols correspond to the stacked spectra: the high-mass stacked spectrum, low-mass stacked spectrum, and full stacked spectrum, respectively. The color bar in the figure indicates metallicities, which were derived by \cite{Bian_2018}.

As shown in Figure~\ref{fig:O32}, the high-redshift galaxies have elevated O32 and R23 values compared with local galaxies. This indicates that galaxies in protoclusters possess higher ionization parameters and lower metallicities. Further analysis reveals that low-mass galaxies display higher R23 values compared to high-mass galaxies, suggesting that low-mass galaxies have lower metallicities.

Additionally, we present a grid spanning a range of ionization parameters and metallicities. This grid demonstrates that with increasing metallicity, the grid shows  the increase of O32 value, whereas the R23 index changes non-monotonically. Due to the non-monotonic behavior of R23, we choose to use O32 instead of R23 for metallicity calibration. The colorbar indicating metallicity values follows the calibration from \cite{Bian_2018}.


\subsection{BPT Diagrams}

\begin{figure*}[htbp]
\centering 
\subfigure{
\includegraphics[width=0.45\linewidth]{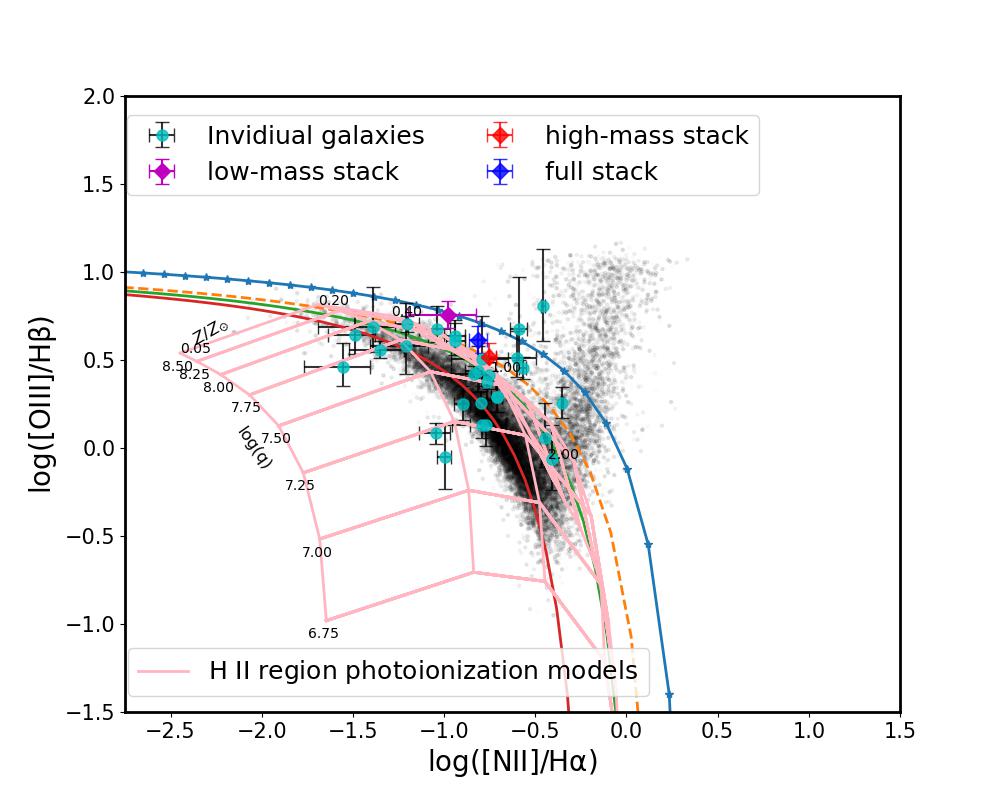}
\includegraphics[width=0.45\linewidth]{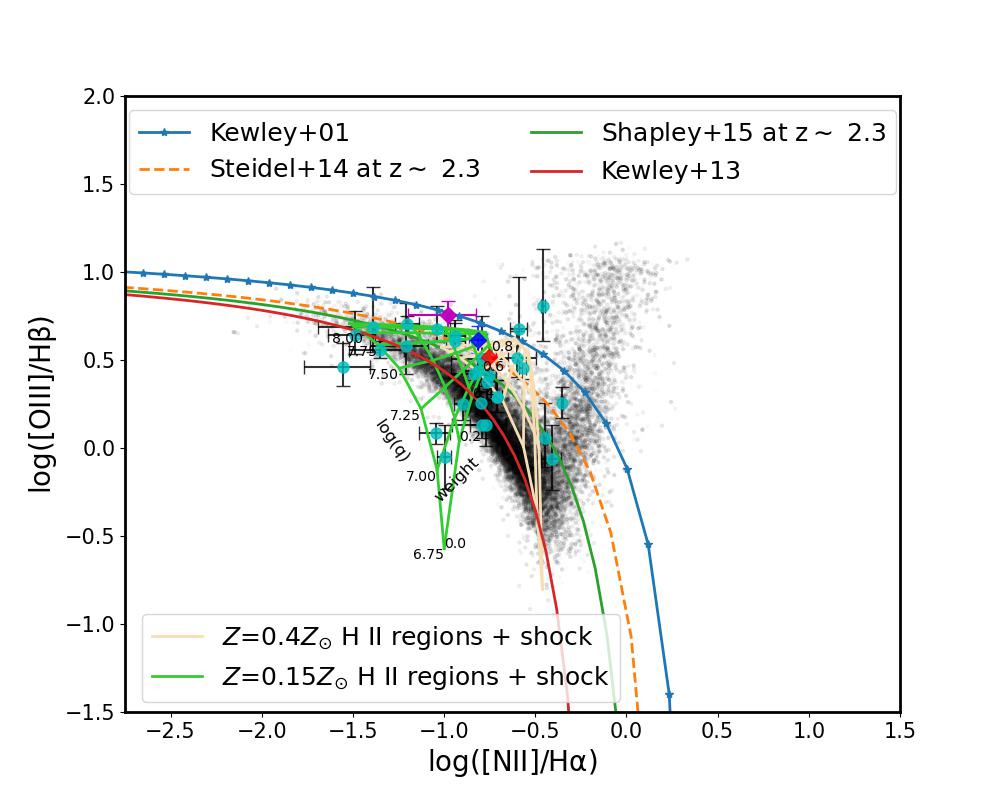}
}
\caption{N2 BPT diagrams for our sample galaxies residing in the cores of the three protocluster fields. Stacked data is from the Table~\ref{table:stacking}, and individual galaxy data is from Table~\ref{table::measure}. The individual galaxy with measurements are indicated by cyan points, and blue point is the full stacked sample by using 43 galaxies in the protocluster fields. Red point is the high-mass stacked sample and purple point is low-mass stacked sample. The grayscale histogram corresponds to the distribution of local SDSS galaxies. The blue curve is the maximum starburst model of \citet{Kewley_2001}. The green line and the orange line both represent observational data, with the green line coming from \citet{Shapley15} and the orange line from \citet{Steidel_2014}. The weight parameter represents the fractional contribution of shocks to the combined model.
{\bf Left}: BPT diagrams with pure \HII region photoionization model (shown in pink).
{\bf Right}: BPT diagrams incorporating the \HII region plus shock-wave photoionization models at two different metallicities: 0.4 $\rm Z_{\odot}$ (shown in wheat) and 0.15 $\rm Z_{\odot}$ (shown in green).
}
\label{fig:N2BPT}
\end{figure*}

\begin{figure*}[htbp]
\centering 
\subfigure{
\includegraphics[width=0.45\linewidth]{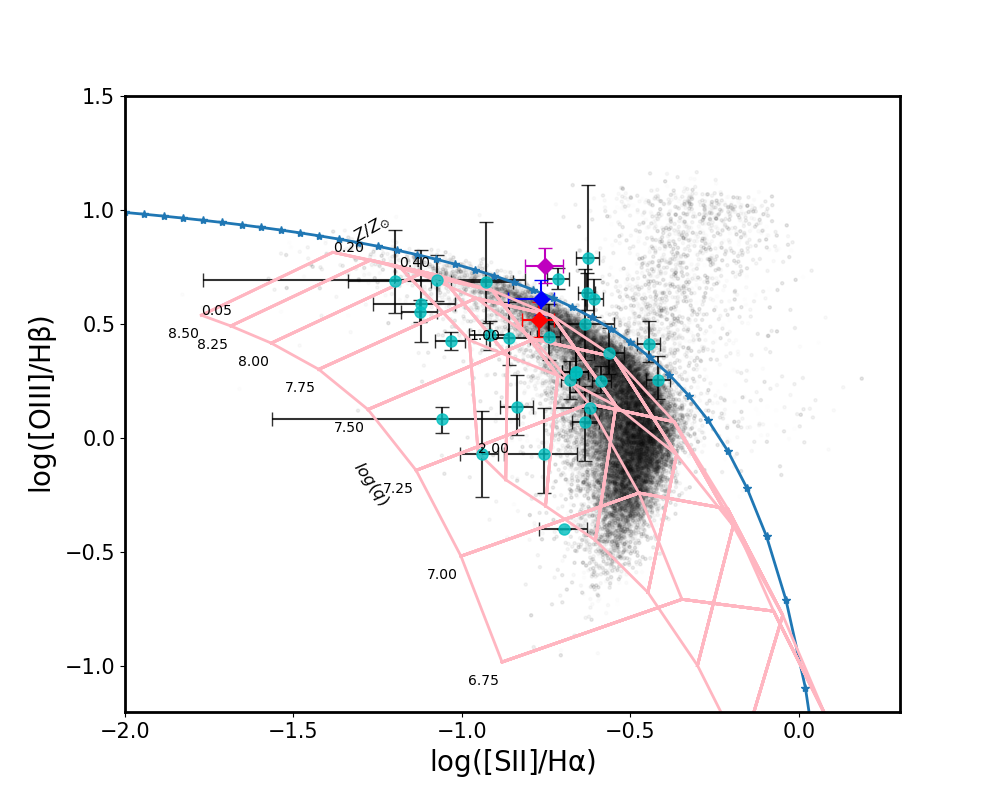}
\includegraphics[width=0.45\linewidth]{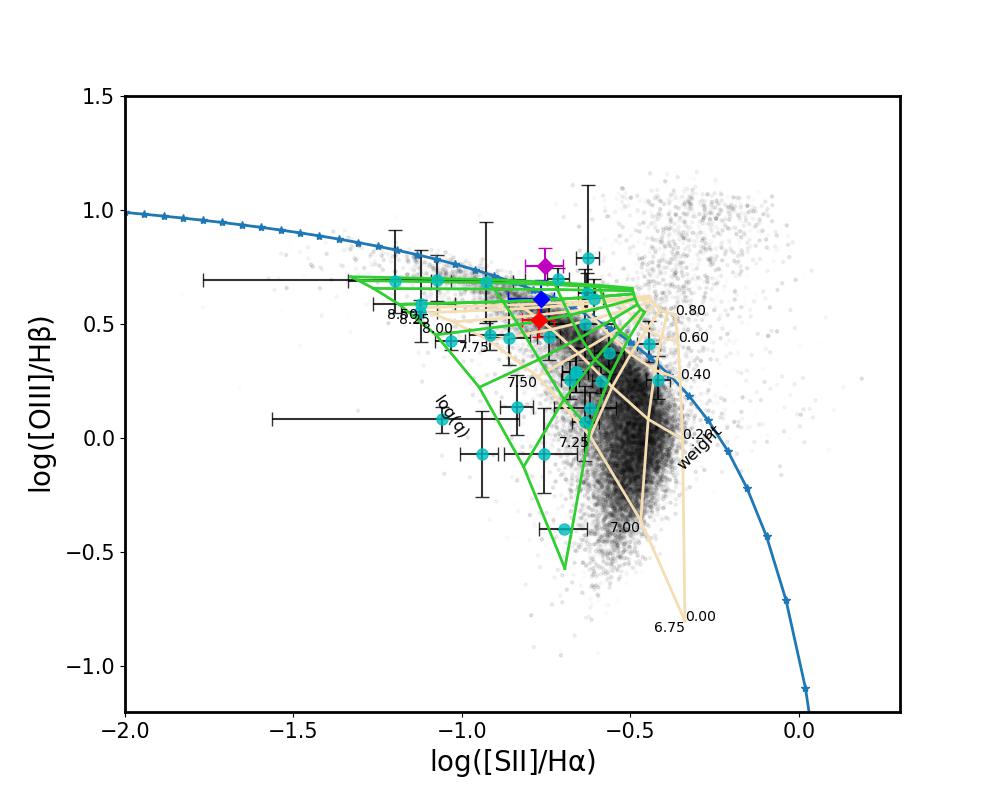}
}
\caption{S2 BPT diagrams for local and our sample galaxies residing in the cores of the three protocluster fields. The color-coding and styles of the symbols and lines follow exactly those shown in Figure~\ref{fig:N2BPT}.
}
\label{fig:S2BPT}
\end{figure*}

The three BPT diagrams are widely used diagnostics for the ionization properties of the ISM in galaxies. All use $\frac{\OIII\lambda5007}{\rm{H}\beta}$ on their Y Axis; each uses a different X axis: $\frac{\NII\lambda6584}{\rm{H}\alpha}$ (N2), $\frac{\SII\lambda\lambda6717,6731}{\rm{H}\alpha}$ (S2), and
$\frac{\OI\lambda6300}{\rm{H}\alpha}$ (O1) 
\citep{,Baldwin1981, Veilleux1987}.

\begin{figure*}[htbp]
\centering 
\subfigure{
\includegraphics[width=0.45\linewidth]{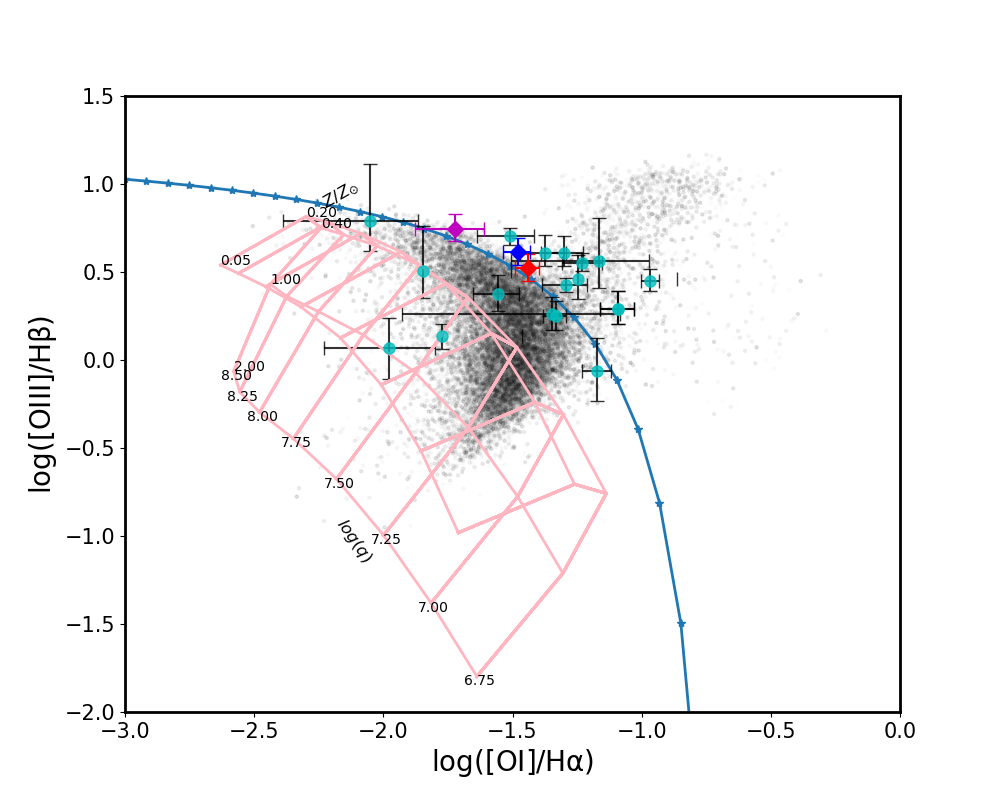}
\includegraphics[width=0.45\linewidth]{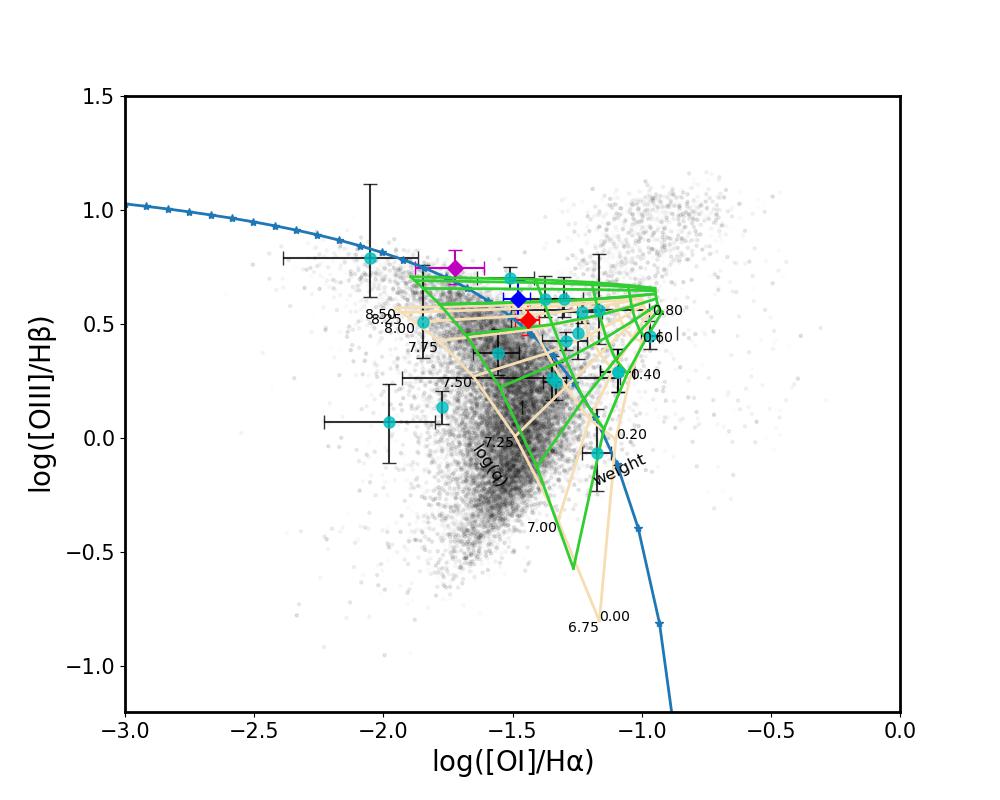}
}
\caption{O1 BPT diagrams for local and our sample galaxies residing in the cores of the three protocluster fields. The color-coding and styles of the sym. bols and lines follow exactly those shown in Figure~\ref{fig:N2BPT}.
}
\label{fig:O1BPT}
\end{figure*}

Figures~\ref{fig:N2BPT}-\ref{fig:O1BPT} present our N2, S2, and O1 BPT diagrams, respectively. The black points represent local galaxies from the SDSS sample, consistent with those shown in Figure~\ref{fig:O32}. The cyan points denote the 43 HAEs confirmed, while the red, purple, and blue rhombic symbols correspond to the stacked spectra for high-mass, low-mass, and total stacked samples, respectively, as in Figure~\ref{fig:O32}. 

The left and right panels of Figure~\ref{fig:N2BPT}-\ref{fig:O1BPT} differ in the grid lines representing two sets of models. The pink grid lines in the left panel correspond to photoionization models considering only \HII regions, while the wheatand green grid lines in the right panel represent models incorporating both \HII regions and shocks with different metallicity. The detailed discussion of these models is provided in Section~\ref{sec:models}.

In the N2 BPT diagram (Figure~\ref{fig:N2BPT}), we find that both the stacked and individual galaxy samples are offset above the local galaxy sequence. Specifically, the low-mass stacked sample exhibits a higher $\OIII\lambda5007/\rm{H}\beta$ ratio, whereas the high-mass stacked sample shows a lower $\OIII\lambda5007/\rm{H}\beta$ ratio. Specifically, the low-mass stacked sample exhibits an offset of 0.45 dex above the local galaxy sequence in the $\OIII\lambda5007/\rm{H}\beta$ ratio, the high-mass stacked sample exhibits 0.36 dex offset above the local galaxy sequence, 
and the full-stacked sample exhibits 0.37 dex offset above the local galaxy sequence.
All stacked samples lie along the sequences defined by \cite{Steidel_2014}--dashed yellow line--and \cite{Shapley15}--solid green line.
Thus protocluster galaxies share the same offset from local galaxies as other high-redshift galaxies at $z \sim 2$.

In contrast, the S2 BPT diagram (Figure~\ref{fig:S2BPT}) reveals almost no significant offset between our samples and the local galaxies. However, in the O1 BPT diagram (Figure~\ref{fig:O1BPT}), the stacked samples lie well beyond the \cite{Kewley_2001} demarcation line, which typically separates star-forming galaxies from AGNs. Notably, the high-mass and total stacked samples exhibit elevated $\OI\lambda6300/\rm{H}\alpha$ ratios, with log($\OI\lambda6300/\rm{H}\alpha$)$\sim$ -1.45. 

These results suggest that the ionization conditions in protocluster galaxies cannot be fully explained by simple stellar photoionization diagnostics alone. 
We therefore conduct a more detailed analysis using more complicated photoionization models.

\section{Modeling the Ionization Mechanisms}
\label{sec:Ionization}

Star-forming galaxies can be distinguished from AGN based on their emission line ratios using the N2, S2, and O1 BPT diagrams \citep{Kewley_2001}. Upon analyzing Figure~\ref{fig:N2BPT} through Figure~\ref{fig:O1BPT}, we observe a distinct offset in our sample compared to local galaxies. Previous studies have shown that high-redshift galaxies exhibit different emission line ratios compared to their local counterparts, complicating the application of these ratios to infer the physical properties of high-redshift galaxies. Several surveys, including the Keck Baryonic Structure Survey \citep[KBSS;][]{kbss,Steidel_2014} and the MOSFIRE Deep Evolution Field Survey \citep[MOSDEF;][]{Shapley15,Sanders_2016}, have reported similar results, highlighting the need for model recalibration when analyzing high-redshift galaxies.

One potential influence is diffuse ionized gas (DIG). However, \cite{Sanders_2017} demonstrate that DIG contributes negligibly in typical high-redshift galaxies, as these galaxies have higher star formation rates \citep{Whitaker_2014} and more compact structures \citep{vanderWel_2014} compared to their local counterparts.

However, it should be noted that our sample galaxies are strong \Ha emitters, indicating relatively high star formation rates. Consequently, the discussion of these sample galaxies may not fully represent all galaxies in protoclusters as a whole.

To further investigate the underlying processes responsible for these offsets in protoclusters, we employ the MAPPINGS software \citep{mapping}, which enables detailed modeling and interpretation of the physical conditions and associated emission line features in these environments. The MAPPINGS V photoionization code calculates ionization and emission properties for elements ranging from hydrogen to zinc, accounting for both recombination and collisionally excited lines across continuum and line fluxes. It is also capable of predicting emission-line spectra of medium- and low-density plasmas subjected to varying degrees of photoionization and shock ionization. MAPPINGS V offers flexibility in key parameters—such as ionization parameter, metallicity, electron density, and shock velocity—enabling simulations to capture the complex conditions in the interstellar medium.

\begin{deluxetable}{lcccccc}
    \label{table:mapping}
    \tablecolumns{7}
    \tablewidth{0pt}
    \tablecaption{Parameter Details for the MAPPINGS Model.}

    \tablehead{
        \HII\ region &  & Z/Z$_{\odot}$ & log($q$) &  \\
        &   & 0.4 or 0.15 & 6.75-8.50  &  
    }
    \startdata
    Slow shock wave\tablenotemark{a} &  $v$(km s$^{-1}$)   &  Z/Z$_{\odot}$  & B($\mu$G)  & \\
    & 200  & 0.4 or 0.15 & 10 &  \\
    \enddata
    \tablenotetext{a}{Shock model includes 50\% contribution from precursor.}
\end{deluxetable}

\subsection{Pure Photoionization Models}
\label{sec:models}

We employ an ionization model to characterize protoclusters, focusing on the physical parameters of \HII regions, including the ionization parameter ($q$), the slope of the ionizing spectrum of stars illuminating the gas, and the electron density ($n_e$). 
Based on the parameters calculated in Section~\ref{sec:measure}, protoclusters exhibit high ionization parameters and low metallicities. The $n_e$ for individual galaxies ranges from 34 to 2500 $\rm cm^{-3}$ , while the $n_e$ for the stacked samples ranges from 93 to 1067 $\rm cm^{-3}$. Therefore, we set the $n_e$ of the model range from 50 to 6500 $\rm cm^{-3}$. The ionization parameter log($q$) for most protocluster galaxies is above 7.5, and for the stacked samples, log($q$) exceeds 7.75. Thus, we assume that protocluster environments have high ionization parameters, with a range for log($q$) set from 6.75 to 8.50. We set the metallicity to 0.15 Z$_{\odot}$ and 0.4 Z$_{\odot}$. The plane-parallel geometry is adopted to simulate \HII regions, with the parameters selected based on \cite{Dopita_2013} and \cite{Allen_2008}, as summarized in Table~\ref{table:mapping}.

It is important to note an apparent discrepancy when comparing Figure~\ref{fig:O32} and Figure.~\ref{fig:N2BPT}. Specifically, the photoionization model grid line corresponding to a high ionization parameter (log($q$) = 7.5) lies below the main locus of the SDSS galaxy sample in the O32 diagram (Figure~\ref{fig:O32}). In contrast, the same grid line appears to traverse the central region of the data distribution in the N2 BPT diagram (Figure~\ref{fig:N2BPT}).
However, this is not a contradiction. The apparent difference arises because the model grids overlaid on the two diagrams are not identical in their metallicity coverage. The \HII region grid in Figure~\ref{fig:O32} spans a wide metallicity range (from Z=2.0 Z$_{\odot}$ to 0.05 Z$_{\odot}$), whereas the grid in Figure.~\ref{fig:N2BPT} is shown for only two discrete metallicity values (0.4 and 0.15 Z$_{\odot}$). Upon closer inspection of Figure~\ref{fig:N2BPT}, the track corresponding to a metallicity of 0.4 Z$_{\odot}$ is indeed positioned consistently with the physical trends and data distribution shown in Figure~\ref{fig:O32}.

Using the Starburst99 \citep{starburst99} spectrum model in MAPPINGS to fit the ionizing flux, we find that the ionizing continuum slope is 21.26.
\subsection{Pure Photoionization Model Results}\label{sec::stack}

The emission line flux ratios of \OIII/\Hb, \OI/\Ha, \NII/\Hb, and \SII/\Ha obtained from the pure photoionization model are shown in the left panels of Figure~\ref{fig:N2BPT} to Figure~\ref{fig:O1BPT}. In the left panels, the green grid lines represent the metallicity model with 0.15 Z$_{\odot}$, while the wheatgrid lines correspond to the model with 0.4 Z$_{\odot}$.

Under these model assumptions, the resulting log(\OI/\Ha) ratios are all above -1.83. However, the observed values for the high-quality galaxies and the overall stacked sample are approximately -1.44. This discrepancy suggests that additional radiation sources may be enhancing the flux of the \OI line. Although we initially attempted to use a pure photoionization model to explain the \OI emission line offset, we found that this approach did not adequately match the observed data. Therefore, additional radiation sources must be considered to explain the offset in the line ratios.

\subsection{Photoionization Combined with Shock Excitation}

Shock excitation can arise from various phenomena, including supernovae, stellar winds, galaxy interactions, and AGN-related activities such as jets \citep{Kewley_2019}. Shocks significantly enhance low-ionization emission lines such as \OI, \NII, and \SII. Since these shocks can originate from different astrophysical processes, quantifying their total contribution poses considerable challenges. Therefore, this paper adopts a weighted summation approach between \HII regions and shocks as a simplified example, demonstrating how combining these two components can improve the overall fit.
\begin{equation}
\label{eq:shock}
\begin{split}
&\rm spectrum_{total} \\
&\rm = (\omega * spectrum_{shock} + (1-\omega)*spectrum_{H_{\scriptsize II}})
\end{split}
\end{equation}
where $\omega$ is the contribution of shocks to the total emission, a fraction between 0 and 1. Higher $\omega$ values indicate a greater influence of shock excitation, while lower values reflect the dominance of photoionization in \HII regions.

The weight $\omega$  represents the contribution of slow shocks. Specifically, the final spectrum is obtained by adding the spectrum of the shocks, scaled by the weight, to the spectrum of the \HII regions. This additional free parameter allows us to simultaneously account for both the ionization driven by star formation and the additional excitation caused by shocks, providing a more comprehensive explanation for the observed emission line ratios in protocluster galaxies. The weight in the figure represents the fractional contribution of shocks to the total emission. Higher weights indicate a more significant role of shock excitation in the emission lines, while lower weights suggest that the emission is predominantly dominated by \HII regions.

Shock waves can also enhance certain emission lines. Slow shocks (velocity $<$ 500 km $\rm s^{-1}$) can enhance emission lines such as \NII and \OI, while fast shocks (velocity $ >$ 500 km $\rm s^{-1}$) can also enhance the \OIII emission line \citep{Kewley_2019}. Given the strong \OI emission lines observed in Figure~\ref{fig:O1BPT}, shock excitation could be a significant contributor to the BPT diagram offsets in our sample. So in the shock model, we adopt a slow shock velocity of 200 km $\rm s^{-1}$, as suggested by \cite{Kewley_2019} and \cite{Cameron_2020}. The metallicity and magnetic field strength are chosen based on the values recommended by \cite{Cameron_2020}. The relevant parameters for the shock model are also summarized in Table~\ref{table:mapping}.
\cite{Allen_2008} reported that magnetic fields influence gas compression. Specifically, in the presence of a magnetic field, the compression is proportional to the Alfv{\'e}n Mach number, which can limit the degree of compression in preshock gas.

We present the mixed model of \HII\ regions and shocks in the right panels of Figure~\ref{fig:N2BPT}-\ref{fig:O1BPT}. The green grid lines represent 0.15 Z$_{\odot}$ , while the wheatgrid lines represent 0.4 Z$_{\odot}$. From the figure, we can observe that comparing two figures, pure \HII region model can not fit the protocluster while shock plus \HII region model could fit the galaxies in protoclusters.

Using the MAPPINGS V software, we find that high-redshift protoclusters are likely affected by shocks, which may account for 0\% to 40\% of the total \Ha emission line flux. This contribution helps explain the observed emission line ratio offsets in protoclusters. As shown in Figure~\ref{fig:N2BPT}, when no shock emission is present, the locus of \HII regions aligns with the distribution of SDSS galaxies, with no offset. However, as the weight of slow shocks increases, log(\NII/\Ha) shifts from the SDSS locus toward the fit lines reported by \citet{Steidel_2014} and \citet{Shapley15}, indicating that shocks may explain the observed offsets in high-redshift galaxies. Similarly, the S2 BPT diagram shows that combining \HII regions and shocks can cover most of the galaxies in our sample. For Figure~\ref{fig:O1BPT}, when there is no shock contribution, the emission line ratios are consistent with local star-forming galaxies. However, as the shock component increases, the ratios gradually move beyond the \cite{Kewley_2001} demarcation line, eventually aligning well with the observed values in protoclusters.

Contrary to \citet{li2025insights} who found that $z > 5$ protocluster galaxies are more metal-rich than field galaxies of equivalent stellar mass, potentially driven by accelerated star formation. However, our overdense sample exhibits the opposite trend: protocluster galaxies show systematically lower metallicities compared to equally massive field galaxies. This discrepancy suggests that our protoclusters at $z\sim2$ are at different evolutionary stages, where different mechanisms may be at play.

\section{DISCUSSION}
\label{sec:discussion}
\subsection{Shock Modeling and the Velocity Dispersion vs. \OI/\Ha\ Diagram}
To distinguish between photoionization from star formation and shock excitation, we constructed a kinematic diagnostic diagram plotting gas velocity dispersion ($\sigma$) against the \OI/\Ha\ emission line ratio.

We generated a grid of theoretical models for comparison using the MAPPINGS V code. The models were computed for two metallicities, $0.15 Z_\odot$ and $0.4 Z_{\odot}$, to bracket the metallicity range of our sample galaxies, which spans from $12 + \log(\text{O/H}) = 8.15$ to $8.6$. A lower metallicity grid is therefore appropriate for our samples. The shock models were generated with parameters consistent with those listed in Table~\ref{table:mapping}, with the only difference being the shock velocity. For a pure photoionization reference, we modeled an \HII\ region with an ionization parameter of $\log(q) = 6.5$. The shock models cover a range of velocities ($v_s = 100, 200, 300, 400$ km $\rm s^{-1}$) to match the observed velocity dispersions, which were derived from the FWHM of the \Ha emission line. The uncertainties on the velocity dispersion for each galaxy were propagated from the measurement errors of the FWHM.

\begin{figure*}[htbp]
\centering 
\subfigure{
\includegraphics[width=0.45\linewidth]{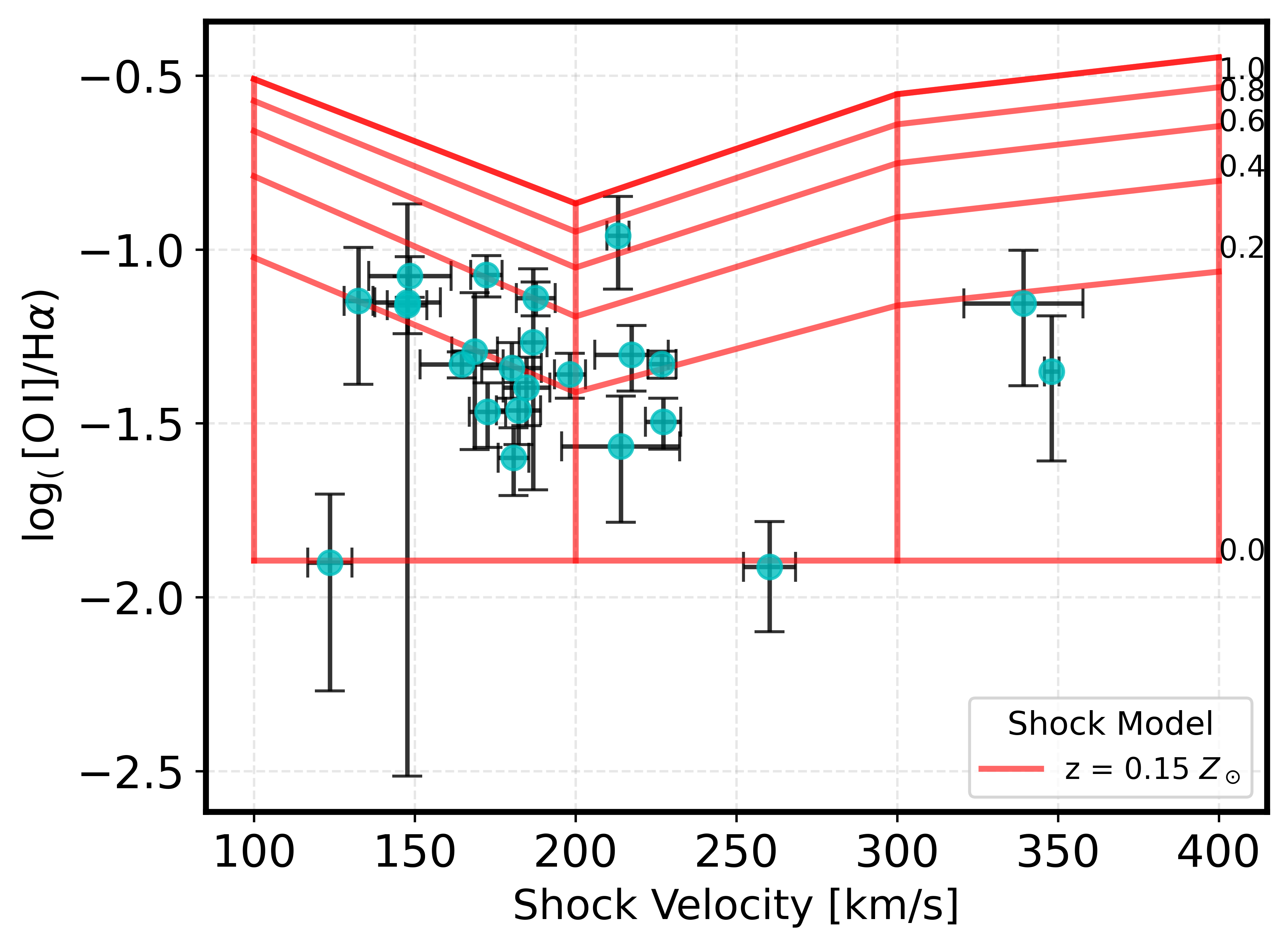}
\includegraphics[width=0.45\linewidth]{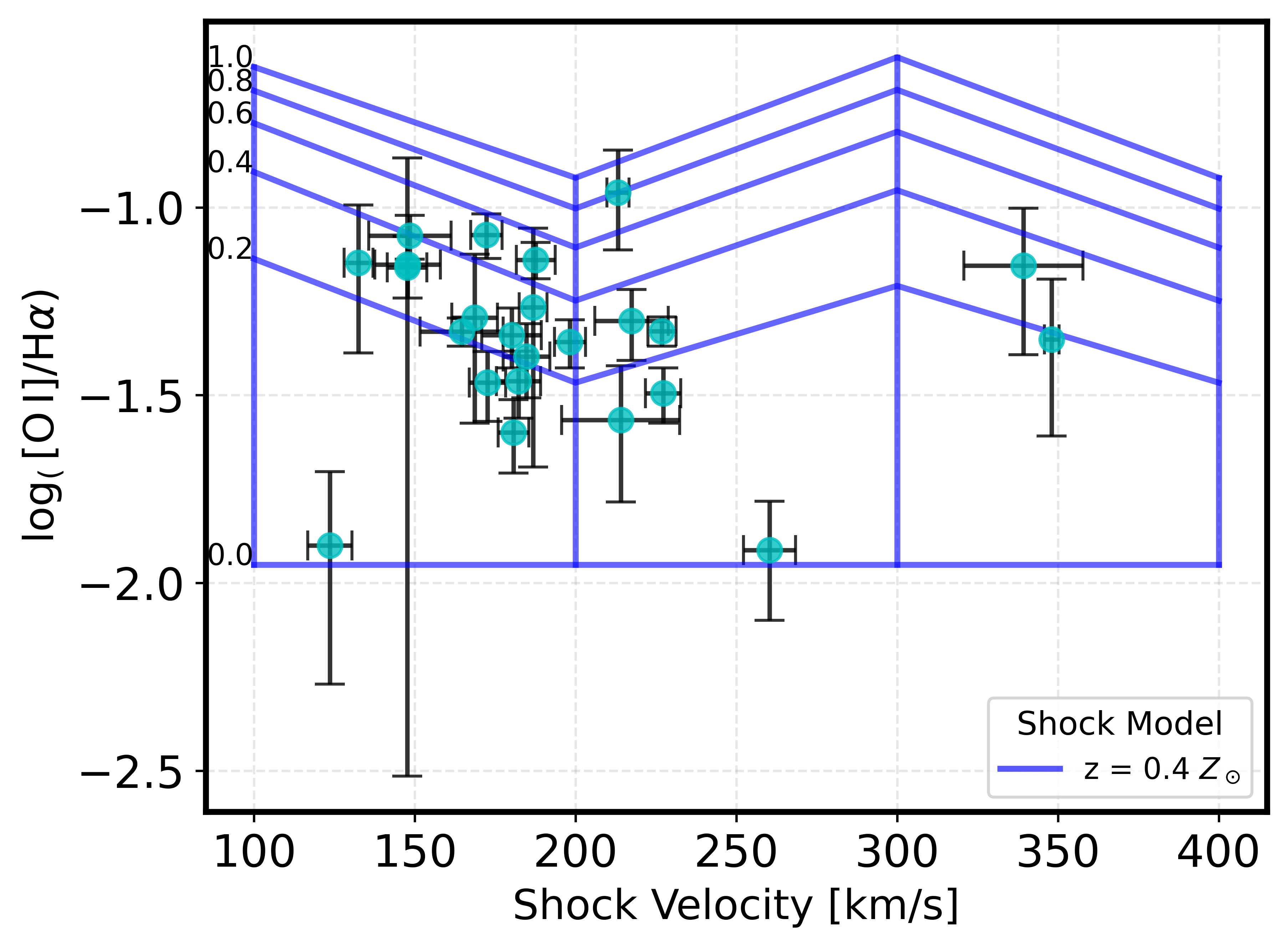}
}
\caption{The velocity dispersion vs. log(\OI/\Ha) diagnostic diagram for our sample galaxies.The cyan data points represent our sample of protocluster galaxies. The solid lines show theoretical model grids from MAPPINGS V for two metallicities: 0.15 $Z_{\odot}$
 (left panel, red) and 0.4 $Z_{\odot}$ (right panel, blue). Within each grid, the sequence of models represents different mixing fractions between a pure \HII region model and a pure shock model. The intermediate lines indicate different shock contributions, ranging from 0\% to 100\%, as indicated by the black labels.}
\label{fig:FWHM}
\end{figure*}

Figure~\ref{fig:FWHM} presents our observational data overlaid on the model grids. The vast majority of our sample galaxies occupy a distinct region of the diagram that cannot be simply explained by pure photoionization. Instead, their positions align remarkably well with the low-metallicity track that represents a combination of an \HII\ region and a shock component.

Notably, most of the data points    are around the parts of the model track corresponding to low-velocity shocks, in the range of 100 $- $ 200 km $\rm s^{-1}$. This provides evidence that slow shocks, rather than high-velocity shocks, are a prevalent mechanism contributing to the gas excitation in this protocluster environment.

The confirmed presence of shocks has significant implications for the determination of gas-phase abundances. Emission from shock-heated gas can contaminate key diagnostic lines, such as \NII, leading to an overestimation of the intrinsic nitrogen abundance. Following the framework established by \cite{flury2024}, we can use our findings to correct for this effect.

Given that most of our galaxies are consistent with shock velocities between 100 and 200 km s$^{-1}$, we adopted a 200 km s$^{-1}$ shock model as a representative case to estimate the shock contribution to the \NII\ flux. We then corrected the observed \NII/\Ha\ ratios that would arise purely from photoionization. Assuming a shock contribution could be as high as 40\%, For our low-mass, high-mass, and total stacked samples, this would lead to a significant decrease in the \NII/\Ha\ ratio, this suppression can be as large as 0.48 dex. This correction brings the stacked line ratios of our sample into agreement with those of local star-forming galaxies from the SDSS. As a result, the inferred nitrogen abundances of our high-redshift protocluster galaxies become consistent with those of their low-redshift counterparts, and are also in agreement with the findings of \cite{flury2024}. It is also a possible resolve result for the paper \citep{Shapley15,Steidel_2014}.

This finding suggests a potential solution to the long-standing puzzle of apparent nitrogen overabundance in high-redshift galaxies \citep{Shapley15,flury2024,Steidel_2014}. Our analysis suggests that this observed excess may not entirely reflect intrinsic differences in chemical evolution, but may be significantly influenced by an observational artifact from unaccounted for shock excitation. This aligns with previous studies by \cite{Newman_2014} and \cite{flury2024}, which also pointed to the potential role of shock excitation mechanisms in biasing abundance measurements.

\subsection{The Environmental Origin of the Observed Shocks}
\label{sec:env_effects}

The environment plays a key role in galaxy evolution. As pioneered by \cite{Dressler1980}, galaxies in high-density regions have undergone a different evolutionary path than their counterparts in the field. 

An important question arises: which environmental mechanisms are responsible for driving the widespread shocks we have detected? In dense protocluster environments, two primary candidates are gravitational interactions, most notably galaxy mergers, and hydrodynamic processes like ram-pressure stripping (RPS).

Galaxy mergers are a well-established mechanism for generating shocks \citep{person24}. Our measured shock contribution to the H$\alpha$ flux (0\%-40\%) and the average gas velocity dispersion (FWHM $\sim 188~\text{km~s}^{-1}$) are both consistent with the properties observed in local merger-induced shock systems \citep{Rich_2011}, making mergers a plausible candidate.

However, to test if mergers are the dominant driver, we must consider their prevalence. Following the morphological classification catalog from \cite{liu_what_2023} for galaxies in these same protocluster fields, we cross-match our sample with catalog from \cite{liu_what_2023}. Out of our 43 galaxies, we find only 7 that exhibit clear merger signatures. While mergers certainly contribute to the overall shock budget, their low observed fraction makes it unlikely that they are the primary mechanism responsible for the widespread shock signatures detected across the majority of our sample. This suggests that a more pervasive mechanism must be at play.

Ram-pressure stripping presents another plausible origin for the shocks. The link between ram pressure and shock formation has been established by work such as \cite{Kang_2011}. Given that RPS is considered a key process in the dense environments of high-redshift protoclusters \citep{Abramson_2014, Strazzullo_2019}, it is a strong candidate for driving the shocks in our sample.

The detection of gas tails is a primary method for identifying the influence of ram-pressure stripping. Historically, RPS was first invoked to explain the morphology of head-tail radio galaxies in nearby clusters \citep{Miley1972, miley1980}. Observationally, these tails can be traced using different gas phases. Tails of neutral atomic gas (\HI) are a classic indicator of RPS. Hot gas tails, visible in X-ray observations, have also been found in massive early-type galaxies. While direct evidence for RPS in molecular gas is still emerging, ionized gas has proven to be an excellent tracer. Numerous one-sided \Ha tails, indicative of RPS, have been observed in nearby clusters like Virgo, Coma, and A1367, and similar evidence has been found in some protoclusters \citep{Strazzullo_2019}. A future study with high-resolution imaging is therefore essential to search for these features and provide a conclusive verdict on the dominant environmental mechanism at play.

In conclusion, our analysis provides evidence that the widespread shocks in these protocluster galaxies are primarily driven that ram-pressure stripping may be the most likely dominant mechanism. This conclusion is supported by two findings: (1) the low observed fraction of major mergers is insufficient to explain the prevalence of shocks, and (2) the galaxies are located in the dense cores of protoclusters, where environmental conditions favor ram-pressure stripping as the primary driver of the observed shocks. A definitive confirmation, however, requires high-resolution imaging to directly detect morphological signatures of these processes, such as stripped gas tails or faint tidal features.

\section{SUMMARY}\label{sec::summary}
In this paper, we present the MAMMOTH-MOSFIRE program, which is designed to obtain deep Keck MOSFIRE K-band spectroscopy of 43 galaxies at $z\sim2.2$. These galaxies are located in the core regions of three most massive protoclusters at cosmic noon (BOSS1244, BOSS1441 and BOSS1542). 

We utilized the PYPEIT software for data reduction, which includes slit tracing, wavelength calibration, and the extraction of 1D spectra from 2D images. During this process, we identified 43 \Ha emitters. For WFC3/G141 grism spectra, we employ the GRIZLI software for reduction.  We also measured stellar masses for all 43 galaxies, finding that 31 are high-mass galaxies with mass range from $\rm 10^{10} M_\odot-10^{11} M_\odot$, while 12 are low-mass galaxies with mass range from $\rm 10^{9} M_\odot-10^{10} M_\odot$.

The electron density in our samples at $z\sim2.2$ was measured using the \SII doublet emission lines, following the method of \cite{Sanders_2016}. As shown in Table~\ref{table:stacking} and Table~\ref{table::measure}, total stacked galaxies have 293 $\rm cm^{-3}$, low-mass stacked galaxies exhibit a relatively high electron density of 1067 cm$^{-3}$, whereas high-mass stacked galaxies show a lower electron density 93 $\rm cm^{-3}$. Individual galaxies span a range from 37 $\rm cm^{-3}$ to 2400 $\rm cm^{-3}$. These measurements provide valuable input parameters for modeling the physical conditions in protoclusters.

Using the calibration described by \cite{Bian_2018}, we find that 80\% galaxies 12+log(O/H) $<$  8.5. Applying the \cite{Kewley_2002} calibration to the O3R2 diagnostic further indicates that about most of the sample lie in the ionization-parameter range log($q$) $\sim 7.5-8.0$. Collectively, these results suggest that protocluster galaxies generally feature high ionization parameters and low metallicities.

In both the N2 and O1 BPT diagrams, we note an offset, suggesting the presence of non-star-forming line emission in the protocluster galaxies. Modeling only \HII regions cannot fully reproduce these offsets, indicating additional ionizing sources. In particular, the strong \OI line is likely driven by shock excitation, prompting us to use the MAPPINGS software to explore these conditions.

By combining emission from \HII regions and slow shocks (velocity $\sim$ 200 km $\rm s^{-1}$), we can model the emission line ratios in the N2, S2, and O1 BPT diagrams.  As summarized in Table~\ref{table:mapping} and shown in the Figure~\ref{fig:N2BPT} and Figure~\ref{fig:O1BPT}, a high ionization parameter log($q$) ranging from 6.75 to 8.50 and a shock contribution of about 0\% to 40\% to the total \Ha flux explain the observed line ratios. Shocks enhance low-ionization lines such as \NII and \OI, shifting the BPT locus to higher ratios.

We conclude that protocluster galaxies at $z\sim2.2$ likely have high ionization parameters, low metallicities, and also experience significant shock excitation. The shocks, combined with \HII region emission, driven by environmental effects, can explain the observed offsets in multiple BPT diagrams. The detection of shocks suggests that ram-pressure stripping may be a primary driver of gas dynamics in these environments, although the role of other mechanisms, such as mergers, cannot be ruled out.
All in all, this work highlights the powerful synergy between high-resolution Keck/MOSFIRE K-band spectroscopy and HST/WFC3 G141 slitless spectroscopy, jointly covering the full suite of rest-frame optical emission lines of galaxies at $z\sim2$.

\vspace{5mm}
\facilities{HST(WFC3), Keck(MOSFIRE)}


\software{PYPEIT \citep{pypeit:joss_arXiv},
          LIME \citep{lime},
          LMFIT \citep{lmfit},
          GRIZLI \citep{grizli}, 
          MAPPINGS \citep{mapping},
          DS9 \citep{ds9},
          SPECTUILS \citep{Specutils},
          }

\begin{acknowledgments}
We thank the anonymous referee for very constructive report that helped improve the quality of this paper, and Xihan Ji for useful discussion.
This work is supported by the China Manned Space Program with grant no. CMS-CSST-2025-A06, the National Natural Science Foundation of China (grant 12373009), the CAS Project for Young Scientists in Basic Research Grant No. YSBR-062, the Fundamental Research Funds for the Central Universities, the Xiaomi Young Talents Program. 
D.D.S. acknowledges support from the National Science Foundation of China (Grant No. 12303015), the National Science Foundation of Jiangsu Province (Grant No. BK20231106).
This work is also supported by NASA through HST grant HST-GO-16276 and HST-GO-17159.
We wish to extend special thanks to those of Hawaiian ancestry on whose sacred mountain we are privileged to be guests. Without their generous hospitality, most of the observations presented herein would not have been possible. This work is also based on observations made with the NASA/ESA Hubble Space Telescope, which is operated by the Association of Universities for Research in Astronomy, Inc., under NASA contract NAS 5-26555.
\end{acknowledgments}

\appendix
\section{Pure \HII Region Photoionization Models for log(\OI/\Ha) Ratio Analysis}

To evaluate the log(\OI/\Ha) ratio, we generated a suite of pure \HII region models with the MAPPINGS code \citep{mapping}, varying key physical parameters. Our protocluster sample spans ionization parameters in the range log($q$) from 7.75 to 8.00, metallicities $12+\log(\mathrm{O/H}) < 8.5$, and electron densities $n_e$ from 37 $\rm cm^{-3}$ to 2400 $\rm cm^{-3}$. By comparing our measured metallicities, ionization parameters, and electron densities with those adopted in the pure \HII region models, we find that the pure \HII region predictions fail to reproduce the elevated log(\OI/\Ha) values observed in our protocluster galaxies. As demonstrated in Section~\ref{sec:discussion}, none of these pure \HII region models can reproduce the elevated log(\OI/\Ha) ratios seen in our protocluster galaxies, indicating that processes beyond pure HII region photoionization is necessary.

\begin{table*}[ht]
\begin{threeparttable}
\centering
\caption{The log(\OI/\Ha) Ratio in \HII\ Regions in Different Physical Parameters.}\label{tab::HIIregion}
\begin{tabular}{c|cccccccc}

\toprule
\toprule
\multicolumn{9}{c}{$n_{e}$ $\sim$ 10 $\rm cm^{-3}$} \\
\hline
\diagbox{Z/Z$_{\odot}$}{log($q$)} & 6.75 & 7.00 & 7.25 & 7.50 & 7.75 & 
8.00 &8.25 & 8.50  \\
\hline
0.05&-1.708&-1.846&-1.983&-2.106&-2.195&-2.248&-2.274&-2.286\\
0.2&-1.264&-1.413&-1.565&-1.708&-1.822&-1.895&-1.934&-1.953\\
0.4&-1.164&-1.323&-1.49&-1.652&-1.786&-1.877&-1.928&-1.952\\
1.0&-1.52&-1.678&-1.855&-2.042&-2.216&-2.346&-2.421&-2.454\\
2.0&-2.354&-2.507&-2.691&-2.887&-3.054&-3.155&-3.183&-3.156\\
\hline
\multicolumn{9}{c}{$n_{e}$ $\sim$ 50 $\rm cm^{-3}$} \\
\hline
\diagbox{Z/Z$_{\odot}$}{log($q$)} & 6.75 & 7.00 & 7.25 & 7.50 & 7.75 & 
8.00 &8.25 & 8.50  \\
\hline

0.05&-1.71&-1.853&-2.002&-2.15&-2.281&-2.379&-2.44&-2.47\\
0.2&-1.265&-1.418&-1.582&-1.749&-1.903&-2.025&-2.107&-2.152\\
0.4&-1.156&-1.32&-1.498&-1.681&-1.852&-1.989&-2.084&-2.138\\
1.0&-1.406&-1.578&-1.767&-1.972&-2.172&-2.338&-2.448&-2.501\\
2.0&-1.909&-2.085&-2.28&-2.48&-2.662&-2.795&-2.861&-2.866\\
\hline
\multicolumn{9}{c}{$n_{e}$ $\sim$ 450 $\rm cm^{-3}$} \\
\hline
\diagbox{Z/Z$_{\odot}$}{log($q$)} & 6.75 & 7.00 & 7.25 & 7.50 & 7.75 & 
8.00 &8.25 & 8.50  \\
\hline

0.05&-1.71&-1.854&-2.008&-2.166&-2.319&-2.455&-2.561&-2.63\\
0.2&-1.26&-1.414&-1.582&-1.758&-1.931&-2.087&-2.212&-2.299\\
0.4&-1.137&-1.303&-1.483&-1.673&-1.859&-2.024&-2.154&-2.241\\
1.0&-1.305&-1.48&-1.671&-1.876&-2.077&-2.251&-2.374&-2.437\\
2.0&-1.638&-1.815&-1.999&-2.181&-2.348&-2.479&-2.556&-2.579\\
\hline
\multicolumn{9}{c}{$n_{e}$ $\sim$ 6500 $\rm cm^{-3}$} \\
\hline
\diagbox{Z/Z$_{\odot}$}{log($q$)} & 6.75 & 7.00 & 7.25 & 7.50 & 7.75 & 
8.00 &8.25 & 8.50  \\
\hline

0.05&-1.706&-1.851&-2.006&-2.168&-2.329&-2.485&-2.624&-2.736\\
0.2&-1.243&-1.397&-1.565&-1.744&-1.925&-2.097&-2.249&-2.371\\
0.4&-1.092&-1.258&-1.44&-1.633&-1.826&-2.003&-2.153&-2.265\\
1.0&-1.158&-1.336&-1.527&-1.731&-1.932&-2.107&-2.235&-2.302\\
2.0&-1.39&-1.567&-1.746&-1.92&-2.079&-2.207&-2.29&-2.323\\
\hline
\multicolumn{9}{c}{$n_{e}$ $\sim$ 30000 $cm^{-3}$} \\
\hline
\diagbox{Z/Z$_{\odot}$}{log($q$)} & 6.75 & 7.00 & 7.25 & 7.50 & 7.75 & 
8.00 &8.25 & 8.50  \\
\hline

0.05&-1.708&-1.852&-2.008&-2.171&-2.336&-2.498&-2.654&-2.794\\
0.2&-1.228&-1.382&-1.55&-1.731&-1.915&-2.093&-2.258&-2.4\\
0.4&-1.042&-1.207&-1.391&-1.59&-1.789&-1.975&-2.135&-2.261\\
1.0&-0.988&-1.17&-1.368&-1.579&-1.786&-1.967&-2.102&-2.179\\
2.0&-1.113&-1.295&-1.479&-1.658&-1.822&-1.955&-2.044&- \\

    \bottomrule
    \end{tabular}
\begin{tablenotes}
      \footnotesize
      \item[a] A dash ($-$) denotes parameter combinations for which no model output is available; these lie outside the density–ionization range relevant to our sample.
    \item[b] All values were computed with the MAPPINGS V photoionization code \citep{mapping}.
    \item[c] In each diagonal header, the left entry is the metallicity $Z/Z_{\odot}$ and the top entry is the ionization parameter $\log (q)$.
    \end{tablenotes}
\end{threeparttable}
\end{table*}


\end{CJK*}

\clearpage
\bibliography{sample631}{}
\bibliographystyle{aasjournal}
\end{document}